\documentclass{emulateapj}
\def\stacksymbols #1#2#3#4{\def\theguybelow{#2}
        \def\verticalposition{\lower#3pt}
        \def\spacingwithinsymbol{\baselineskip0pt\lineskip#4pt}
        \mathrel{\mathpalette\intermediary#1}}
\def\intermediary #1#2{\verticalposition\vbox{\spacingwithinsymbol
        \everycr={}\tabskip0pt
        \halign{$\mathsurround0pt#1\hfil##\hfil$\crcr#2\crcr
                \theguybelow\crcr}}}
\def\lta{\stacksymbols{<}{\sim}{2.5}{.2}}
\def\gta{\stacksymbols{>}{\sim}{3}{.5}}

\shorttitle{SYNCHROTRON EMISSION FROM CYGNUS A}

\begin{document}


\title{First Dynamic 
Computations of Synchrotron Emission from the Cygnus 
A Radio Cavity; Evidence for Electron Pair Plasma in Cavity}
\author{William G. Mathews\altaffilmark{1}}

\altaffiltext{1}{University of California Observatories/Lick
Observatory,
Department of Astronomy and Astrophysics,
University of California, Santa Cruz, CA 95064
mathews@ucolick.org}

\begin{abstract}
Cosmic rays, thermal gas and magnetic fields in FRII radio cavities
are assumed to come entirely from winds flowing from just behind the
jet shocks.  Combining analytic and computational methods, it is shown
that the computed radio-electron energy distribution and synchrotron
emissivity spectra everywhere in the Cygnus A radio cavity agrees with
radio observations of the Cygnus A lobes.  The magnetic field energy
density is small everywhere and evolves passively in the post-shock
wind.  Most synchrotron emission arises in recent post-shock material
as it flows back along the radio cavity wall.  Because it experienced
less adiabatic expansion, the magnetic field in this young backflow is
larger than elsewhere in the radio lobe, explaining the observed radio
synchrotron limb-brightening.  
The boundary backflow decelerates due to small
cavity pressure gradients, causing large-scale fields perpendicular to
the backflow (and synchrotron emission) to grow exponentially unlike
observations.  However, if the field is random on subgrid (sub-kpc)
scales, the computed field reproduces both the magnitude and slowly
decreasing radio synchrotron emissivity observed along the backflow.
The radio synchrotron spectrum and image computed with a small-scale
random field agree with VLA observations.  The total relativistic
energy density in the post-jet shock region required in computations
to inflate the radio cavity matches the energy density of 
relativistic electrons observed in the post-shock 
region of Cygnus A.  This indicates that the component in the jet and
cavity that dominates the dynamical evolution is a relativistic pair
plasma.

\end{abstract}

\vskip.1in

\section{Introduction}

Our recent 2D gas 
dynamical calculations of the remarkable FRII radio-X-ray 
source Cygnus A explicitly 
include the dynamical effects of both hot 
thermal gas and relativistic gas inside the radio cavity
(Mathews \& Guo 2010, 2012; MG10 and MG12 hereafter). 
Since the dynamics of 
relativistic, radio-synchrotron-emitting electrons 
are known, we now extend the results of MG12 
by computing the radio synchrotron emission
and magnetic flux throughout the radio cavity and 
compare with detailed VLA observations of Cygnus A. 

Unlike all previous FRII computations,
we do not explicitly compute the jet evolution, 
but instead assume that the energy that inflates the radio cavity
enters via the well-observed post-jet shock region.
It is assumed that all thermal, non-thermal and magnetic 
energy inside the radio cavity originates in this small
region just behind the jet shock.
This intensely energetic and possibly 
turbulent region is the source of a powerful
wind that flows away in every direction except, we assume,
upstream in the jet direction.
With these assumptions, it is possible to 
compute the entire evolution of Cygnus A with a post-shock 
source region that moves out into the cluster gas;
the jet is not explicitly computed.
Avoiding the jets 
is convenient since sufficiently detailed 
observations of faint FRII jets do not currently exist
while observations of the physical properties of the 
luminous post-shock regions are much better understood. 

One of the chronic computational difficulties experienced 
by previous attempts to compute FRII dynamics, 
like those of Cygnus A, 
is the Kelvin-Helmholtz 
instability (KHI) that invariably occurs when gas, 
after leaving the jet post-shock region, 
is diverted to flow back in the anti-jet direction 
toward the cluster center as it inflates the radio cavity.
The large velocity shear generated by this backflowing gas 
gives rise to KHIs that cause the computed cavity to become 
irregular in shape and to produce chaotic, irregular 
internal velocity fields.
Neither of these attributes are observed in Cygnus A 
or are common in other FRII sources.
Many or most FRII radio cavities have 
smoothly curved outer boundaries and observed radio-synchrotron 
ages that decrease in a smooth, monotonically fashion along the 
jet direction, showing no evidence that the interiors of 
radio cavities are scrambled by large scale KH-induced 
turbulence although turbulence on 
smaller scales is still possible. 
In particular, VLA observations of the radio cavity of Cygnus A 
reveal a very smooth, gently curved boundary that 
appears to be stable to KHI.

Computations driven by a moving post-shock source region 
allow us to reproduce a number of important radio and X-ray 
features observed inside the Cygnus A radio cavity.
One of our most important results in MG12 is the realization that 
relativistic gas backflowing from the jet-shock 
is restricted to a rather 
narrow shell just adjacent to the outer boundary of the 
radio cavity.
Since this flow carries with it the youngest, most energetic 
radio-emitting electrons and the largest fields, 
we conclude that 
radio synchrotron emission from the cavity arises mostly 
or exclusively from this boundary backflow.
Although this radiating shell geometry has not been considered by 
most observers who interpret radio or inverse Compton 
X-ray emission from the radio lobe, 
edge-brightened radio synchrotron 
emission has indeed been verified in 
Cygnus A by Carvalho et al. (2005) and in other 
FRII sources by Daly et al. (2010).

The KHI in the boundary backflow must be damped.
This can be accomplished 
either with a magnetic field along
the flow direction or by adding a small viscosity.
In MH12 we adopted viscous damping 
and this choice is further supported in the discussion 
in Section 2 below.
Viscous damping arises naturally from the entrainment of
a small amount of cluster gas into the rapid backflow. 
Entrainment also decelerates 
the low-inertia, thermally relativistic 
backflow, but most of the backflow deceleration is due to 
a small negative radial pressure gradient inside the cavity.
(The gradient is small because of the large scale height 
of hot cavity gas in the cluster gravity field.)
In turn, the deceleration significantly amplifies 
large scale magnetic 
fields in the backflow that are not parallel to 
the backflow velocity. 

Observations of Cygnus A indicate that magnetic energy densities
throughout 
the cavity region are less than the combined energy densities 
of local thermal and non-thermal gas components,  
and this is the assumption made in MG12.
Consequently, the Lorentz term ${\bf j \times B}$ 
does not appear in the equation of motion, i.e. the 
large scale magnetic field is not strong enough to 
influence the flow of gas or relativistic plasma, 
and the frozen-in magnetic field evolves passively as it 
flows from the post-shock region into the radio cavity.
This passivity allows us to explore the effects of 
various field morphologies in the boundary backflow 
using a post-processing procedure based on the 
same gasdynamical evolution described in MG12.

The radio-synchrotron emission throughout the radio cavity 
in Cygnus A is completely 
determined by the energy distribution of 
relativistic synchrotron electrons 
observed in the post-shock region 
$n(\gamma)$ between energy $\gamma = E/m_e c^2$
and $\gamma + d\gamma$.
In Section 3 we describe how simple analytic solutions 
can be used to translate this post-shock energy spectrum 
to an evolved spectrum anywhere anywhere in the lobe 
and in particular along the boundary backflow.

Another new result discussed in MG12 is that the 
small regions of luminous radio and X-ray emission near 
the tips of the jets, often referred to as ``hotspots'', 
are more complex than previously 
recognized.
These kpc-sized regions 
of intense radio synchrotron emission in FRII sources are not 
in general located at the energetic jet-shock source regions 
described above, but in arc-shaped regions just ahead 
along the jet direction where the post-shock wind first 
crashes against dense (shocked) cluster gas and is 
powerfully compressed. 
In Cygnus A
radio and synchrotron-self Compton X-ray 
emission just behind the jet shock is much weaker 
than from the bright arc about 1.5 kpc ahead. 
Nevertheless, the post-shock region is certainly 
the hottest spot in the radio cavity, 
i.e. the hottest spot is not the observed ``hotspot''. 
Fortunately, in at least one very powerful FRII source it is 
possible to observe X-radiation from both spot regions
(Erlund et al. 2010), so a new terminology is needed to distinguish 
between the post-shock 
and compressed arc-shaped regions. 
VLA observers describe significant radio structure in and  
near the hotspots (e.g. Black et al. 1992 and 
Leahy et al. 1997) and occasionally suggest
a spatial distinction between the jet termination shock 
and bright radio emission, 
but it is unclear how these features relate to 
the spatial distinction we discuss here 
for which we provide a dynamical model.

In MG12 we chose to refer to the (typically fainter) post-shock
region as the ``hotspot'' and the observationally brightest 
region as the ``offset''.
Here we drop the use of ``hotspot'' altogether and 
unambiguously refer 
to the post-shock region as the ``shock spot'' and to the 
more luminous arc-shaped wind compression offset ahead 
as the ``bright spot''. 

Detailed hydrodynamic modeling for Cygnus A 
such as we describe in MG12 and discuss here at length, 
requires considerable computing and analysis
and therefore must be restricted to 
a small  set of well-chosen 
parameters such as the core-hotspot distance, the age,
the magnetic field, jet luminosity, 
initial properties of the cluster gas in which the 
event occurred, etc. 
Among the best known parameters is 
the physical size of Cygnus A from core to hotspots,
determined from its redshift 
with $H_0 = 75$ km s${-1}$ Mpc$^{-1}$, is about 
60 kpc with 1 kpc per arcsecond (e.g. Wilson, Smith \& Young 2006).
We adopt an age of 10 Myrs for Cygnus A 
which can be determined from 
the spectral evolution of the radio continuum 
(e.g. Machalski, Chyzy, Stawarz \& Koziel 2007)
who find 10.4 $\pm$ 1.6 Myrs using equipartition fields 
or from the size and strength 
of the bow shock 
and a simple self-similar evolutionary model in the 
cluster atmosphere 
($\sim 3$ Myrs from Wilson, Smith \& Young 2006).
Other previous estimates of the age do not differ significantly
from our chosen value, 10 Myrs 
(e.g. Begelman \& Cioffi 1989; Carilli et al. 1991; 
Kaiser \* Alexander 1999).
The average rate that cosmic ray energy is supplied to 
the radio cavity by the jet is approximately 
$L_{cr} = 10^{46}$ ergs s$^{-1}$ 
(e.g. Wilson, Smith \& Young 2006).
Our assumed value of $L_{cr}$ refers only to the 
single radio lobe 
we calculate, so the total power is $L_{tot} = 2L_{cr}$.
However, this power is sufficient to inflate radio 
cavities to volumes only about half that 
observed in each Cygnus A cavity after 10 Myrs. 
Consequently, in \S6 below 
we consider a second hydrodynamic model with 
$L_{cr} = 2.65 \times 10^{46}$ ergs s$^{-1}$
that does reproduce the observed cavity volume.
Magnetic field strengths 
measured by combining radio and X-ray 
observations are independent of equipartition assumptions 
and expected to be reasonably accurate, 
but they are computed with simpler cavity 
emission geometries than we describe here. 
Yaji et al. (2010) determine a (presumably uniform) 
field of 20 $\mu$G in the Cygnus A radio cavity by comparing 
radio synchrotron emission with inverse Compton X-ray 
emission upscattered from the microwave background and 
synchrotron-self Compton (SSC) X-rays produced by 
upscattered synchrotron emission. 
Their analysis (especially for SSC) 
depends on the geometric morphology of 
the synchrotron-emitting region 
which Yaji et al. (2010) assume is uniform throughout 
the radio lobe.
This differs considerably from the edge-brightened 
radio lobe emission we describe in MG12. 
Nevertheless, until a new data analysis is performed 
using our model,
we adopt 20 $\mu$G as a reference ``observed''
value for the radio lobe field. 
Our adopted field in the bright spot, about 220$\mu$G, 
taken from 
Stawarz et al. (2007), is less dependent on global 
dynamical models. 
Overall, uncertainties in these key Cygnus A parameters, 
and the dynamical models we compute from them,
are likely to be 
dominated more by systematic errors due to inadequate 
underlying assumptions than by statistical observational
errors.

\section{Summary of Computed Dynamical Models for 
Cygnus A}

Our computations of radio synchrotron emission 
from Cygnus A rely largely on hydrodyamic models 
discussed in MG10 and MG12. 
For this reason it is useful to briefly summarize the 
assumptions and results in those papers.

We consider the self-consistent dynamics 
of a two-component fluid: thermally (but not kinematically)
relativistic cosmic rays (CRs) and hot gas 
having energy densities $e_c$ and $e$ respectively. 
The pressures of these two fluids are related to the 
energy densities by $P_c = (\gamma_c - 1)e_c$ 
and $P = (\gamma - 1)e$ respectively where 
$\gamma_c = 4/3$ and $\gamma = 5/3$. 
CR pressure gradients communicate momentum to the gas 
by means of small magnetic fields frozen into the gas
that are otherwise dynamically insignificant; 
Alfv\'en speeds are generally small 
compared to typical gas velocities. 
The magnetic energy density $u_B = B^2/8\pi$ 
inferred from radio and X-ray 
observations of the Cygnus A radio lobes 
is smaller than $e_c$ by factors of 10-600 
(Hardcastle \& Croston 2010; Yaji et al. 2010). 
Observed fields in the radio lobe are 
small, only $15-20\mu$G (Yaji et al. 2010). 
Even the much larger magnetic fields 
observed in Cygnus A bright spots, $\sim 220\mu$G, 
indicate that $u_B$ is several times smaller than $e_{c,rad}$, 
the energy density of synchrotron-radiating CR electrons alone  
(Stawarz et al. 2007).
Consequently, for the approximate computations discussed 
here we ignore the Lorentz force 
${\bf j \times B}$ on the gas and 
regard the magnetic field as passively moving with the 
hot gas velocity.

The equations we consider are:
\begin{equation}
{ \partial \rho \over \partial t}
+ {\bf \nabla}\cdot\rho{\bf u} = {\dot \rho}_{ss}
\end{equation}
\begin{equation}
\rho \left( { \partial {\bf u} \over \partial t}
+ ({\bf u \cdot \nabla}){\bf u}\right) =
- {\bf \nabla}(P + P_c) 
+ {\bf \nabla\cdot\Pi }
- \rho {\bf g} + \rho {\bf a}_{ss}
\end{equation}
\begin{equation}
{\partial e \over \partial t}
+ {\bf \nabla \cdot u}e = - P({\bf \nabla\cdot u})
+ {\bf \Pi : \nabla u}
\end{equation}
\begin{equation}
{\partial e_c \over \partial t}
+ {\bf \nabla \cdot u}e_c = - P_c({\bf \nabla\cdot u})
+ {\dot S}_{ss}
\end{equation}
\vskip.1in
\begin{equation}
{ \partial \tau \over \partial t}
+ {\bf \nabla}\cdot\tau{\bf u} = 0.
\end{equation}
\vskip.1in
\begin{equation}
{\partial {\bf B} \over \partial t} = 
{\bf \nabla \times (u \times B)}
\end{equation}
\begin{equation}
{\bf \nabla \cdot B} = 0.
\end{equation}
These equations are solved in 2D axisymmetric 
cylindrical coordinates, 
appropriate for the very symmetric X-ray image 
of Cygnus A (MG12).

In equation (3) we omit a term for radiative losses from 
the thermal gas component -- this is justified by the 
remarkably short age of the Cygnus A event,
only 10 Myrs.
Equation (4) for the integrated 
CR energy density does not include 
CR diffusion nor do we include loss terms due to synchrotron or 
inverse Compton emission, since 
strongly radiating CRs are a small
fraction of the total CR energy density required to 
inflate Cygnus A.
A mass conservation equation for the
relativistic CR particles is unnecessary 
because of their negligible rest mass. 

The viscous stress tensor ${\bf \Pi}$ appearing in both the momentum 
and internal energy equations is proportional to 
the (assumed constant and isotropic) viscosity $\mu$. 
These terms are provided 
in cylindrical coordinates in the Appendix of MG12.
We assume a classical Navier-Stokes form for the viscous terms, 
but it must be emphasized that 
the physical nature of 
(turbulent or particle) transport processes 
in relativistic, weakly magnetic plasmas 
are very poorly understood  
(Schekochihin et al. 2010). 

In our 2D computations, the gasdynamical evolution of Cygnus A
is entirely driven by an outflowing wind 
from the moving shock spot source region immediately behind 
the termination shock of the jet.
The mean velocity of the shock spot is determined by its 
distance from the cluster core 
and the approximate age of Cygnus A, 
$v_{ss} = 60 {\rm kpc}/10{\rm Myrs} = 5870~{\rm km~s}^{-1}$.
For simplicity we assume this mean velocity is also the 
instantaneous velocity (e.g. O'Dea et al. 2009). 
The shock spot velocity is maintained by an acceleration 
source term $\rho {\bf a}_{ss}$ in equation 2,
see MG12 for details.
Our Cygnus A computations are the first to adopt 
shock spot-driven gasdynamics; 
the jet is not explicitly included in the calculation 
but its momentum and energy determine those of the shock spot.
Shock spot-driven flows are preferred because the physical 
properties of the much fainter jets remain uncertain 
while flow variables in the shock spot can be found from 
those observed in the relatively nearby bright spot
(Section 3.3).

All magnetic field and cosmic rays inside the radio cavity 
are assumed to originate in the shock spot wind. 
The cosmic ray source term in equation (4) 
${\dot S}_{ss} = L_{cr}/V_{ss}$ where 
$V_{ss} = 4 \times 10^{64}$ cm$^{3}$ is the shock spot volume 
and $L_{cr} = 10^{46}$ erg s$^{-1}$ is the 
approximate energy supply required to 
inflate the radio cavity to its present volume.
After each time step the magnetic field in the shock spot 
is reset to a fixed value $B_{ss}$ consistent with synchrotron 
self-compton (SSC) X-ray observations 
in the nearby bright spot region (Stawarz et al. 2007).
The assumed constancy of $B_{ss}$ during the Cygnus A 
evolution will need to be verified by future FRII observations.
The shock spot wind is also assumed to 
continuously provide a small 
admixture of non-relativistic plasma onto which the 
magnetic field can be frozen, as represented 
by the ${\dot \rho}_{ss}$ term in equation (1). 
Non-relativistic gas is thought to arrive in the 
jet, entrained from low-entropy cluster gas 
that flows up along the Cygnus A symmetry axis (MG12). 
The amount of non-relativistic gas entering the shock spot
is very uncertain since it cannot be directly observed.

Equation (5) describes the advection of 
$\tau \equiv \rho t_{exss}$,
the product of the local gas density and the exit time 
from the shock spot (in Myrs).
At any point inside the cavity
the time at which the local gas exited the shock spot
can be found from $t_{exit}(z,r) = \tau/\rho$. 
The age of cosmic ray electrons 
at every grid zone inside the cavity 
is $t_{age}(z,r) = 10 - \tau/\rho$ Myr 
where 10 Myr is the current time.

The powerful shock spot wind moves out in all directions, 
but we assume it is unable to move upstream into the 
oncoming jet. 
By this means, material in the shock spot is endowed 
with a net forward momentum acquired from the incident jet.
The subsequent flow of the shock spot wind, confined 
and shaped 
by the dense wall of (shocked) cluster gas on all sides, 
moves to the cavity surface then back along the surface at 
high velocity. 
The shock spot wind transforms into a ``boundary backflow'' 
that moves in the anti-jet direction just along the 
outer boundary of the radio cavity.
Most of the radio synchrotron emission in the Cygnus A cavity 
comes from recently shock spot-energized electrons  
in this boundary backflow.
The observed radio synchrotron 
emission is in fact mostly confined to this backflowing shell 
(Carvalho et al. 2005).
The backflow rapidly decelerates 
due to the small negative radial pressure gradient 
$dP/dz$ inside the cavity. 
At a distance $z = 55$ kpc from Cygnus A center, 
about 5 kpc back from the bright spot,
the backflow velocity is $9 \times 10^4$ km s$^{-1}$, 
but the velocity drops to $\sim 2 \times 10^4$ km s$^{-1}$ 
at $z = 40$ kpc and decreases further beyond 
(see Tables 1 and 2 below).

The high velocity of the boundary backflow relative 
to the surrounding gas generates a strong KH instability
that has appeared in all previous FRII calculations 
(e.g. Hodges-Kluck \& Reynolds 2011 and
Huarte-Espinosa et al. 2011).
The KH instability eventually disrupts the surface of 
the radio cavity in ways that are inconsistent with 
many FRII radio images like Cygnus A which have 
radio cavities with relatively smooth outer boundaries.
Even more troublesome are radio observations of the 
approximate 
spectral ages of synchrotron electrons which vary 
smoothly and monotonically, increasing along the backflow 
without the age mixing expected from KHI.

It is clear that the KH instability must be suppressed 
and this can be done most simply either with relatively strong 
magnetic fields along 
the backflow or with viscosity which we prefer.
Figure 1 shows a superposition of many velocity vectors 
from MG12 for Cygnus A at time 10 Myrs 
without viscosity (top) and with 
a small viscosity (bottom), $\mu = 30$ gm cm$^{-1}$ s$^{-1}$.
The first onset of the KH instability indicates 
which surface of the boundary backflow is unstable,
inner or outer.
The top image in Figure 1 reveals that the KH-unstable backflow 
near $z \approx 34$, $r \approx 9$ kpc is 
initially diverted inward toward the symmetry axis of the radio cavity.
The anti-clockwise motion that results is in the same sense
as the shear at the inner boundary, but opposite to 
the shear at the outer boundary. 
Evidently,
the backflow first becomes KH unstable 
at its inner surface well inside the radio cavity.
(Most previous discussions have assumed that the KH instability
occurs at the radio cavity wall.)

As in Figure 1,
all computed axisymmetric images of the Cygnus A radio cavity in MG12 
are shaped like rockets with sharply pointed leading edges.
However, radio observations 
of Cygnus A (e.g. Fig. 1 of MG12) and most or all other FRII 
radio lobes show a much more rounded blimp-shaped leading edge.
As we discuss in MG12, this difference is 
probably due to the rapid and 
frequent random re-direction of the Cygnus A jet through small angles.
Such deflections can arise during interactions with 
transverse gas density gradients in the filament of thermal cluster 
gas that moves along the the same jet axis. 
In our axisymmetric calculations a single active shock spot 
is constrained to move 
without deflection right along the symmetry axis 
($z$ direction).
By comparison, in Cygnus A a multitude of simultaneously active 
shock spots located near the leading cavity boundary,
each drive individual   
boundary backflows that converge into a single flow 
as they proceed back toward the cluster core.
This results in a much more rounded cavity than those generated 
by a single shock spot.
Radio cavity widths are therefore expected to exceed 
those in our axisymmetric calculations, 
particularly near the leading edge.

Figure 2 shows radio polarization observations 
at 43 GHz of the  
luminous northwestern bright 
spot A in Cygnus A at 0.2\arcsec resolution 
(Carilli et al. 1999).
The unknown field morphology arriving in the jet 
provides the seed for the post-shock field 
which is compressed in the shock, becoming more 
aligned parallel to the shock surface and 
perpendicular to the jet direction.  
The post-shock field may be more disturbed 
than it appears and could be amplified by turbulence. 
A similar polarization pattern is observed 
in the southeastern bright spot D.
The field alignment in bright spots 
could be consistent with a largely toroidal field, 
although substantial radial field components $B_r$ are also likely.
However, 
Figure 3 shows that the polarization of the eastern radio cavity 
in Cygnus A is far from toroidal downstream from the shock spot.
Deviations from pure axisymmetry 
can convert an initially toroidal field into poloidal, 
so the complex radio cavity field in Figure 3 can 
in principle have evolved 
from shock spots having (what appear to be) more ordered fields. 
Another source of non-axisymmetry are the frequent changes 
in the jet direction, as evidenced by multiple bright spots 
in Cygnus A and other FRIIs (see MG12 for a discussion). 
In any case, for simplicity and other computational 
reasons discussed below, 
in our gasdynamical calculations 
we initially consider only toroidal fields in the shock spot which,
due to ideal axisymmetry,  
remain toroidal as they evolve downstream in the cavity.

The induction equation (6) for a toroidal field $B \equiv B_{\phi}$
\begin{equation}
{\partial B \over \partial t} =
-{\partial \over \partial z}(B u_z)
-{\partial \over \partial r}(B u_r)
\end{equation}
automatically satisfies the solonoidal condition 
${\bf \nabla \cdot B  = 0}$.
This equation is linear in $B$ 
so the computed passive field can be rescaled to any desired 
initial value in the shock spot $B_{ss}$ 
as long as the field energy is not dynamically important.
According to equation (8)
an initially uniform backflow field 
($\partial B/\partial z = 0$) grows exponentially
$B \propto \exp(|\partial u_z/\partial z|t)$
as it decelerates in the $z$-direction 
along the backflow ($\partial u_z /\partial z < 0$). 
Evidently, a radial field component $B_r$ would
increase in a similar fashion, but $B_z$ would be 
much less affected by deceleration in the $z$-direction.

To stabilize KHI in the backflow with magnetic forces, it is 
necessary that large scale fields in the direction 
of the backflow $B_z$ provide a Lorentz force 
${\bf j \times B}$ that 
resists deformation at the shear interface.
Toroidal fields have no effect on the KH instability.
While the fields observed in Cygnus A are far too 
small to provide KH stabilization, it is useful 
to consider this case anyway considering the 
uncertainties involved. 
When the field is aligned with the flow,
KH stabilization requires that the 
mean Alfv\'en speed $v_A = B/(4 \pi \rho)^{1/2}$ across 
an interface exceed the change in the shearing 
flow velocity at the interface.
As discussed in MG12,
the Alfv\'en speed is much less than the computed 
radio cavity shear flows when $v_A$ is evaluated with 
$20\mu$G, the radio lobe field observed by 
Yaji et al. (2010), and the local density 
of non-relativistic gas $\rho$ in our computed flows. 
To stabilize the KHI, either $B_z$ would need to be about 10 times 
larger than $20\mu$G or the gas density $\rho$ would need 
to be about 100 times smaller 
(by reducing the rate of gas supply to the shock spot).
However, if $\rho$ is much lower than the value 
adopted by MG12, the velocity of the shock spot wind and 
the backflow would increase considerably due to 
a much lower inertia.
This would cause the apparent ages of synchrotron-emitting 
electrons in the Cygnus A backflow 
to appear younger than observed. 
These arguments, together with the apparent absence of 
large scale $B_z$ fields inside the radio cavity, 
sharply reduce the 
likelihood that KH can be successfully stabilized 
by magnetic fields\footnotemark[2].

\footnotetext[2]{
Shock spot-driven FRII evolution introduces significant 
numerical challenges when poloidal fields
($B_z$ and $B_r$) are included in the shock spot source.
Most or all available standard MHD codes adopt the 
"constrained transport'' procedure in solving the 
induction equation (6).
With constrained transport 
if the initial field is divergence-free, 
the computed field is guaranteed to remain divergence-free 
at later times to machine accuracy.
For shock spot-driven hydrodynamics the difficulty 
maintaining ${\bf \nabla \cdot B} = 0$ arises 
when assumed shock spot fields move forward on the grid 
or need to be reset to constant or 
other time-varying values after each time step. 
This reset breaks the field continuity at the shock spot 
boundary, producing magnetic monopoles and non-zero 
${\bf \nabla \cdot B}$.
When monopoles are present the computed field may look 
physically acceptable and yet be significantly in error.
One solution for this, enthusiastically recommended 
for users of the FLASH3 code, is to spatially diffuse non-zero 
${\bf \nabla \cdot B}$ throughout the 
entire grid, evidently minimizing the influence of 
magnetic monopoles at any particular 
location.
}

We prefer KH stabilization 
with viscosity which contributes to the deceleration 
of the boundary backflow.
Non-thermal jets moving through cluster gas
appear to be decelerated by the entrainment of cluster gas 
(e.g. Bicknell 1994; Laing \& Bridle 2002).
Like the Cygnus A backflow, outward moving  
jets with internal energies dominated by relativistic particles 
have high energy density but very low momentum.
Jet momentum is very sensitive to the decelerating inertia 
provided by small amounts of 
relatively stationary thermal gas that 
enters across FRI jet boundaries. 
Evidently entrainment can arise from a turbulent boundary 
layer at the jet-cluster interface, but the cores of jets 
also decelerate, implying that the entrained cluster gas 
continues to reduce momentum by viscosity 
as it diffuses toward the jet core.
Viscosity must accompany entrainment.
But the deceleration of the Cygnus A backflow differs 
from that in outflowing FRI jets 
and in fact most of the backflow deceleration 
is due to a small negative 
pressure gradient inside the cavity. 
(This is consistent with the finding in MG12 that the 
contribution of viscosity to backflow deceleration 
is small.)
While the transport of non-relativistic 
cluster gas into jets or the Cygnus A boundary 
backflow may be due to complex 
plasma microinstabilities or turbulent activity,
our computations adopt a standard Navier-Stokes 
formalism with uniform viscosity which we assume is adequate for now.

Some of the viscosity experienced by the computed backflow 
may have a numerical origin caused by the 
(unavoidable) numerical entrainment 
of cluster gas as the boundary backflow 
flows at an angle across the computational grid. 
Numerical entrainment may result from mixing and averaging gas  
between adjacent grid zones when the density gradient 
is not perpendicular to the grid zone boundary. 
This numerical effect may mimic the physical entrainment
process and viscous deceleration that occurs in observed jets. 
However, in our computations numerical viscosity 
alone was insufficient 
to suppress KHI. 
To accomplish this it was necessary to explicitly include  
non-zero viscous terms in equations (2) and (3).
Nevertheless, the higher density 
of entrained non-relativistic gas (and associated viscosity)
near the cluster gas-backflow boundary helps to 
stabilize the backflow-cluster gas surface against KHI.
This gradient of entrained cluster gas 
across the boundary backflow 
may explain why the KH instability in our calculation 
first appears in the internal boundary of the flow,
not at the radio cavity-cluster gas interface.

\section{Evaluation of the energy distribution and 
synchrotron emission from cosmic ray electrons}

\subsection{Evolution of the energy distribution}

The number density of cosmic ray electrons 
$n(\gamma,{\bf r},t)$ with energy 
between $\gamma = E/m_e c^2$ 
and $\gamma + d\gamma$ evolves according to
\begin{equation}
{ \partial n \over \partial t}
+ {\bf \nabla}\cdot n{\bf u} =
{ \partial  \over \partial \gamma}
(-{\dot \gamma}n ).
\end{equation}
The energy $\gamma$ decreases by expansion and
by synchrotron plus inverse Compton (IC) losses
\begin{equation}
{\dot \gamma} = -\left({\gamma \over t}  + {\gamma^2 \over
  t_*}\right).
\end{equation}
The first term on the right expresses the assumption
that the expansion can be regarded as spatially uniform 
on small spatial scales.
Local uniform expansion is plausible because 
the spatial density gradients ${\bf \nabla}n$ are small 
in the Cygnus A cavity and 
both the shock spot wind and subsequent backflow velocities 
are subsonic.
Consequently, we assume that the second term on the left 
in equation (9) can be approximated as
${\bf \nabla}\cdot n{\bf u} \approx n{\bf \nabla}\cdot {\bf u}$
and for uniform expansion the velocity divergence is simply
${\bf \nabla}\cdot {\bf u} = 3/t$.

In the second term of equation (10) 
$t_*$ is the characteristic energy loss time 
due to synchrotron emission in a magnetic field with
energy density $u_B = B^2/8\pi$ and inverse Compton 
(IC) losses by scattering cosmic microwave background 
(CMB) radiation
$u_{cmb} = a[T_{cmb}(1+z)]^4$, where $T_{cmb}(1+z)$ is the 
redshifted radiation temperature and 
$a$ is the radiation constant.
The rate of energy loss of each electron depends on 
its pitch angle $(B\sin\theta)^2/8\pi$ 
so the pitch angle distribution must be known or assumed to 
compute the collective energy loss rate.
Kardashev (1962) and Pacholczyk (1970) assume that the initial pitch 
angle distribution at time $t_0$ is isotropic and that each electron 
maintains this same pitch angle $\theta$ during its later evolution. 
However, 
most modern authors follow Jaffe \& Perola (1973) 
and assume that the pitch angle distribution 
is continuously randomized by 
(poorly understood) wave-particle interactions 
on time scales that are shorter than other times of interest.
It is implicitly assumed that pitch angles can be randomized 
without affecting $n(\gamma)$ or without introducing 
appreciable dissipative heating. 
In view of its physical plausibility 
we also adopt continuous pitch angle istropization where 
$\langle (\sin\theta)^2 \rangle = 2/3$ and the 
$\theta$-averaged value of $t_*$ is
\begin{equation}
{1 \over t_*} = {4 \over 3}{\sigma_T \over m_e c}
\left( u_{cmb} + {2\over 3}{B^2 \over 8 \pi}\right).
\end{equation}

Ideally, when evaluating $t_*$ in a particular 
hydrodynamic grid zone in the radio cavity,
the appropriate magnetic field 
is the time-averaged field experienced by electrons 
in this zone during their passage from the shock spot to the 
zone $\langle B \rangle_{path}$. 
This field could be determined by computing the 
mean field from a number of passive Lagrangian 
particles introduced in the shock spot. 
However, we avoid this complication here 
by simply adopting the local field in the backflow, 
$B \approx \langle B \rangle_{path}$. 
This approximation is expected to be quite good because 
the deceleration along the boundary backflow 
ensures that the mean time-averaged 
field seen by synchrotron-emitting 
electrons is dominated by the current field at any time.

With these simplifying assumptions 
the particle energy distribution equation (9)
can be written as 
\begin{equation}
{ \partial n \over \partial t}
- \left( {\gamma \over t} + {\gamma^2 \over t_*}\right)
{\partial n \over \partial \gamma}
= 2\left({\gamma \over t_*} - {1 \over t}   \right)n.
\end{equation}
This first order partial differential equation has been solved 
by integrating along its characteristic trajectories with solution
\begin{equation}
n(\gamma,t) = K \gamma^{-p}
\left( t \over t_0\right)^{-(2+p)}
\end{equation}
\begin{displaymath}
~~~~~~~~\times
\left[1 - \left( \gamma t_0 \over t_* \right)
\left( t \over t_0 \right)
\ln\left( t \over t_0\right)  \right]^{p-2}
\end{displaymath}
This particular solution\footnotemark[3] is designed to match 
an initial condition with 
a single initial power law distribution 
$n(\gamma,t_0) = K \gamma^{-p}$.
The evolution of $n(\gamma,t)$ 
due to expansion is represented by the 
first term and the second term describes 
the effects of radiative losses.
The second term increases with time 
until $n(\gamma,t) = 0$ when
\begin{equation}
{\gamma t_0 \over t_*} \cdot {t \over t_0} \ln\left({t \over t_0}
\right) > 1.
\end{equation}

\footnotetext[3]{
This solution differs from that published by Kardashev (1962)
in which
$(t/t_0)\ln{(t/t_0)}$ is replaced with $(t/t_0) - 1$ 
where $t_0$ is the time when the electrons first appear behind 
the shock.
As $t \rightarrow t_0$ (and $(t/t_0)\ln(t/t_0) \rightarrow 0$)
$n(\gamma,t) \rightarrow n(\gamma,t_0)$. Times $t < t_0$
are irrelevant since they precede the emission history 
of the electrons in a given grid cell.
}

To adapt this solution to our computed gasdynamical models for 
Cygnus A, we translate the time variation in the expansion 
term in $n(\gamma,t)$ using the relation
$\rho \propto t^{-3}$ for uniform expansion. 
Decreases in the gas density convert directly to 
a measure of time elapsed during uniform expansion 
in equation (13).
Furthermore, $e_c \propto \rho^{4/3}$ 
so the time $t/t_0$ in the expansion part of the solution can
be replaced with $(e_c/e_{c,ss})^{-1/4}$, 
\begin{equation}
n(\gamma,t) = K \gamma^{-p}
\left( {e_c \over e_{c,ss}}\right)^{(2+p)/4}
\end{equation}
\begin{displaymath}
~~~~~~~~\times \left[1 - \left( \gamma t_0 \over t_* \right)
\left( t \over t_0 \right)
\ln\left( t \over t_0 \right)  \right]^{p-2}.
\end{displaymath}
Here $e_c$ is the current CR energy density in any 
cavity grid zone  
and $e_{c,ss}$ is the energy density in the shock spot 
at the retarded time when the gas in this grid zone left 
the shock spot.

Figure 4 shows the variation of the mean energy density 
in the shock spot with computation time 
$\langle e_{c,ss}(t)\rangle$. 
By computation time we mean the time since the beginning 
of the Cygnus A event. 
Although the rate that cosmic ray energy is supplied 
to the moving shock spot ${\dot S}_{ss}$ remains constant 
in our dynamical calculations, 
the energy density $e_{c,ss}(t)$ decreases with time with the 
local cluster gas density. 
At early times, 
when the local cluster gas pressure is high, the 
energy density ${\dot S}_{ss}\Delta t$ supplied 
to the shock spot in timestep 
$\Delta t$ cannot flow as rapidly away from the shock spot
and the resulting $e_{c,ss}$ is larger.
At early times 
relativistic gas in the shock spot is inertially confined 
by the higher density and pressure 
in the ambient cluster gas.
The solid line fit in Figure 4 is 
$\langle  e_{c,ss}(t)\rangle = 1.85\times 10^{-8} 
+ 8.69\times 10^{-10}(10 - t)^{1.9}$ erg cm$^{-3}$
where time is in Myrs.
Likewise, although non-relativistic gas is continuously 
added to the shock spot at a uniform rate, the gas density 
in the shock spot decreases with time for the same reasons.
The scatter in Figure 4 
is due to numerical transients
as the shock spot moves from one grid zone to the next. 

The time factor in the radiative loss 
term $t/t_0$ of equation (15) 
is the ratio of the age of the cosmic rays in each grid 
zone (time since leaving the shock spot) divided by the 
computation time when the cosmic rays left the shock spot 
$t_{exss} = \tau/\rho$,
\begin{equation}
{t \over t_0} = {t_{age} \over t_{exss}}
= { 10 - t_{exss} \over t_{exss}}.
\end{equation}
The computation age of Cygnus A is currently 10 Myrs.

\subsection{Evaluation of the local synchrotron emissivity 
spectrum}

The synchrotron emissivity (erg cm$^{-3}$ s$^{-1}$) 
is given most simply by
\begin{equation}
\epsilon_{\nu} d \nu = - {dE \over dt} n(\gamma)
{d\gamma \over d\nu} d\nu
\end{equation}
where $E = \gamma m c^2$ and $n(\gamma)$ is the number 
density of emitting
electrons between $\gamma$ and $\gamma+d\gamma$.
The power emitted by a single electron 
\begin{equation}
{dE \over dt} = -{4 \over 3} \sigma_T \beta^2 \gamma^2 c u_B
\end{equation}
is assumed to be concentrated at a single frequency
\begin{equation}
\nu \approx \gamma^2 \nu_g ~~~{\rm where}~~~ \nu_g
= \left({e \over 2 \pi m c}\right)B
\end{equation}
is the non-relativistic gyrofrequency of an electron in 
the local field $B$.
Then
\begin{equation}
{d\gamma \over d\nu} = {1 \over 2 (\nu \nu_g)^{1/2}}.
\end{equation}
Assuming $\beta = 1$ and combining these factors,
the synchrotron spectrum is 
\begin{equation}
\epsilon_{\nu} = {4 \over 3} \sigma_T \gamma^2 c u_B
\cdot n(\gamma,t) \cdot {1 \over 2 (\nu \nu_g)^{1/2}}
\end{equation}
where 
\begin{equation}
\gamma = \gamma(\nu) = \left({\nu \over \nu_g} \right)^{1/2}.
\end{equation}

\subsection{Initial conditions in the shock spot}

Stawarz et al. (2007) determined the 
energy spectrum of synchrotron-emitting cosmic ray 
electrons in the Cygnus A bright spots where the postshock 
wind first impacts the dense wall of (shocked) cluster gas ahead. 
(In Stawarz et al. (2007) 
the ``bright spots'' were referred to as ``hotspots''.)
The particle spectrum in the bright spots is described
with a double power law with no detectable low energy cutoff:
\begin{equation}
n(\gamma) = \min[n_1(\gamma),n_2(\gamma)]
\end{equation}
where
\begin{equation}
n_1(\gamma) = K \gamma^{-p_1} ~~~{\rm for}~~~ \gamma < \gamma_{cr}
\end{equation}
\begin{equation}
n_2(\gamma) = K \gamma_{cr}^{p_2 - p_1} \gamma^{-p_2}
~~~{\rm for}~~~ \gamma > \gamma_{cr}
\end{equation}
where $p_1 = 1.5$ and $p_2 = 3.3$ and $\gamma_{cr} = 2000$
are almost the same for Cygnus A bright spots A and D and the 
average value of $K$ is $K_{bs} = 1.1 \times 10^{-4}$ cm$^{-3}$.

Because of the spatial offset of the  
bright spot from the shock spot, the amplitude $K = K_{bs}$ of 
the particle energy distribution and the field strength $B_{bs}$ 
observed in the bright spot must be corrected to estimate shock spot 
values, $K_{ss}$ and $B_{ss}$.
This correction can be done using our dynamical model for Cygnus A 
(MG12) as shown in Figure 5. 
In making this correction, 
we consider volume-averaged corrections for the two shock spot 
zones and two similar adjacent zones 
(in the $r$-direction) located at the bright spot. 
Figure 5 however only shows the correction for the innermost zone.
Finally, we assume that the shape of the
observed double power law is unchanged
during its rapid flow from the shock spot to the offset bright
spot ($\sim 10^4$ yrs); this assumption is discussed further below.

$K$ scales with the total cosmic ray energy density,
$K_{ss} = (e_{c,ss}/e_{c,bs})K_{bs} = 1.38K_{bs} = 1.49\times
10^{-4}$, using computed profiles from MG12 and 
$K_{bs} = 1.1\times                        
10^{-4}$ from Stawarz et al. (2007). 
To map the evolution of cosmic rays from the shock spot to 
any grid zone in the radio cavity, 
$K$ in equations (24) and (25) must be replaced 
with $K_{ss}$. 
Similarly, 
the mean field estimated by Stawarz et al. (2007) in regions A and D,
$B_{bs} = 220\mu$G, must be lowered to 
obtain the true shock spot value,
$B_{ss} = 0.28B_{bs} = 62\mu$G, 
again using MG12 gasdynamical results as shown for the innermost 
zone in Figure 5.
The compressed field in the bright spot is significantly larger 
than that in the shock spot.


As the initial double power law energy distribution 
in the shock spot evolves as it flows into the radio cavity 
along the boundary backflow, the two 
power law cosmic ray populations 
both vary according to equation (13) and  
continue to intersect so 
\begin{equation}
n(\gamma,t) = \min[n_1(p_1,\gamma,t),n_2(p_2,\gamma,t)]
\end{equation}
for $t > t_0$ and with $K = K_{ss}$ and $B = B_{ss}$. 

\section{Radio synchrotron emission from toroidal field}

In this first test of the procedure described above 
we assume, as in MG12,  that the magnetic field in the shock spot, 
and therefore in the radio cavity downstream, is purely toroidal.
The objective is to reproduce the observed  
emissivity profiles (transverse to the 
Cygnus A symmetry axis) $\epsilon_{\nu}(r)$ 
at $\nu = 1.345$ GHz shown with dashed lines in Figure 6.
These emissivity profiles across the full diameter of 
Cygnus A were found by Carvahlo et al. (2005)
by cylindrical deconvolution of the flux profiles
at each crossectional cut shown with solid lines in Figure 6. 
Each panel in Figure 6 is labeled with the distance measured 
in kpc from the Eastern bright spot 
$D_{bs} = 60 - z$ kpc.
The emissivity scale 
(in units of $9.0\times 10^{-34}$ erg cm$^{-3}$ Hz$^{-1}$) 
is shown on the right in Figure 6.
The spatial resolution of the VLA in Figure 6 
is 1.36\arcsec at $\nu = 1.345$ GHz which, at the 
convenient distance of Cygnus A, is about 1.36 kpc, 
comparable to about two computational grid zones in MG12.

Figure 7 shows computed emissivity profiles 
$\epsilon_{\nu}(r)$ at 
$\nu = 1.345$ GHz in the spatially offset bright spot 
"bs'' at $z_{bs} = 60$ kpc and inside the radio cavity 
at $z = 50$, 40, 30, 20 and 10 kpc 
distant from the center of Cygnus A.
The three open circles in Figure 7 are approximate 
peak observed emissivities 
from Figure 6 at $z = 60$, 50 and 40 kpc.

The observed peak emissivities at 50 and 40 kpc 
are displaced from the computed peaks in both the 
horizontal and vertical directions.
The horizontal shift indicates that the observed 
width of the Cygnus A radio cavity 
at these large values of $z$ is somewhat larger 
than those in our gasdynamical model.
Clearly, it would be easy to alter the width of 
the computed cavity by allowing the cosmic ray 
luminosity in the shock spot zones 
${\dot S}_{ss}$ to increase with time.
More likely, the computed width of the radio cavity near 
the shock spot is small because the shock spot 
is constrained to move exactly along the $z$-axis. 
By comparison, as discussed in MG12, 
the jets in Cygnus A 
undergo abrupt changes in direction near the 
symmetry axis, causing the shock spot energy to be 
distributed more broadly than in our computation, 
i.e. redirected jets and their shock spots broaden 
the front surface of the cavity.
For these reasons the horizontal shifts  
of peak $1.345$ GHz emissivity in Figure 7 are not  
a serious problem, easily corrected with a 
more detailed 3D computation.

More important is the emissivity mismatch 
in the vertical direction in Figure 7.
Computed peak emissivities across the boundary backflow 
are 10-30 times larger than those observed 
in $40 \lta z \lta 50$ kpc.
In addition, the computed 
backflow emissivity at $\nu = 1.345$ GHz 
increases systematically from $z = 50$ to 30 kpc,  
a trend noticed already in MG12,
while the observed peak radio emissivity slowly decreases 
throughout this region (see Fig. 6).
The increase in computed $\epsilon_{\nu}$ is largely 
due to the exponentially increasing toroidal magnetic field 
that accompanies backflow deceleration, 
$B \propto \exp (|\partial u_z/\partial z|t)$ 
(see MG12 for details).
However, there is no observational 
evidence of magnetic field variation 
along the Cygnus A radio cavity (e.g. Yaji et al. 2010).

Figure 8 shows the detailed energy spectrum 
$n(\gamma)$ and emissivity $\epsilon_{\nu}$ 
in the shock spot (solid curves) and 
at four distances along the backflow.
The downstream profiles labeled $k = 1-4$ show 
$n(\gamma)$ and $\epsilon_{\nu}$
at the point of maximum 1.345 GHz emissivity  
further along the backflow.
Table 1 lists the spatial coordinates of these positions 
($r,z$) where the 1.345 GHz emissivity peaks for each value of $z$. 
Also listed are the field strength, emissivity 
and backflow velocity all evaluated at $r,z$. 
The right and left limits on $n(\gamma)$ in Figure 8 
vary to match the fixed frequency range in the 
lower panel, but $n(\gamma)$ can be extrapolated beyond 
these ragged limits.
While $n(\gamma)$ decreases as expected 
along the boundary backflow,
the peak emissivity $\epsilon_{\nu}$ is non-monotonic. 
After leaving the shock spot the emissivity  
at $\log \nu = \log (1.345~{\rm GHz}) = 0.13$ 
increases from $k = 1$ to 3
due to the increasing magnetic field.
Profiles labeled $k = 4$ show effects of radiative losses.

In summary, the synchrotron emission from 
a purely toroidal field does not provide a good match
to observation.
The computed field is everywhere much larger than the radio 
lobe field $20\mu$G observed by Yaji et al. (2010). 
The computed 
maximum magnetic field in the backflow $\sim 200\mu$G 
is comparable to the local energy density $e_c$, violating 
our assumption of small Lorentz forces.
Moreover, it is unlikely that this model can be saved by 
including poloidal field components in the shock spot 
source, assuming that the computational monopole problem 
discussed in footnote 1 could be 
overcome. 
Both radial $B_r$ and toroidal $B_{\phi}$ field components 
would be exponentially 
amplified in the decelecrating backflow.
To avoid this amplification, these field components 
in the shock and bright spots 
would need to be converted almost entirely to 
a strong field along the backflow direction $B_z$.

\section{Radio synchrotron emission from random field}

We now depart from the usual MHD procedure, in which the 
induction equation 6 is solved 
for the evolution of the magnetic field. 
Traditional 
MHD hydrocodes can only capture fields with structure 
no smaller than the computational grid.
However, it is useful to consider  
fields that are random on scales smaller than the grid size, 
possibly due to turbulence or poorly understood 
plasma microinstabilities inside the radio cavity. 
Flux conservation requires that the energy density of 
a random magnetic field advects like relativistic fluid, 
$u_B \propto \rho^{\gamma_B}$ with $\gamma_B = 4/3$.
The usual 
frozen-in condition expressed by the induction equation 6 
requires that the field follow the flow velocity of  
non-relativistic gas inside the cavity.
By contrast, a sub-grid random field follows 
the gas density, $B_{ran} \propto \rho^{2/3}$.
\footnotemark[4]

\footnotetext[4]{
In the limit of very high computational grid resolution,
not considered here,  
it should be possible to capture random fields 
using the induction equation and 
reproduce the $B_{ran} \propto \rho^{2/3}$ 
dependence that we consider here. 
}

A subgrid turbulent or otherwise tangled field 
has a number of attractive features.
It is consistent with our assumption of 
continuous pitch angle isotropization. 
It is also consistent with KH damping by quasi-isotropic 
viscosity and cluster gas entrainment 
which imply that the small scale kinematics 
of momentum-carrying non-relativistic 
gas is random at some level. 
Numerical simulations of MHD turbulence in single phase 
fluids often indicate that an initially weak 
field grows in strength until it 
saturates at a value comparable with 
the thermal kinetic energy, but many details remain obscure  
(e.g. Brandenburg, Sokoloff \& Subramanian 2012).
Numerical simulations of turbulence in a two-fluid mixture 
of relativistic and non-relativistic gases (with $e_c \gta e$), 
the case of interest here, have not been performed 
to our knowledge. 
The relatively small
field energy observed in the Cygnus A cavity
suggests that turbulent 
saturation may occur when $u_B$ approaches the thermal 
energy density $e$ which is less than the total CR 
energy density $\sim e_c$. 
Under these circumstances the magnetic field would have 
sub-equipartition values in agreement with fields 
observed in many FRII lobes including Cygnus A.

Guided by these imponderables, we assume that 
the random, small-scale field inside the Cygnus A radio cavity 
at any position and time can be found from the local gas 
density,
\begin{equation}
B_{ran} = B_{ss}(\rho / \langle \rho_{ss}\rangle)^{2/3}
\end{equation}
Where $B_{ss} = 62\mu$G as determined above\footnotemark[5] 
and $\langle \rho_{ss}(t)\rangle$ is the shock spot gas density 
at the retarded time when it first emerged from the shock spot.
The time dependence of $\langle \rho_{ss}(t)\rangle$ 
is illustrated in Figure 9. 
Equation (27) is an idealized model that may not 
apply precisely to the compressed 
and aligned field in the shock spot.
When a field described by equation (27) 
undergoes a one-dimensional compression, 
the field could in principle appear similar 
to the perpendicular field in Figure 2, 
but factors of order unity, not considered here, 
may be required 
to reconcile $B_{ss}$ with the fully random 
field in the backflow.

\footnotetext[5]{
Equation (27) cannot be used to modify 
the field observed in the bright spot to
find the shock spot field as we did with the
toroidal field assumption.
The problem is that equation (27)
predicts that the two spots have about the same
radio synchrotron emissivity which is inconsistent with observations 
of Cygnus A and other FRII sources.
In view of this we imagine that quasi-toroidal fields
parallel to the shock surface 
do exist in both spots as before,
so that the toroidal corrected shock spot value of 
$B_{ss} = 62\mu$G also holds when the random field 
approximation (eqn. 27) applies further downstream. 
This is reasonable because the flow time from shock 
spot to bright spot ($\sim 10^4$ yrs) is very short and the field 
morphology may not change much.
Therefore we assume that the field
experienced by most radio-emitting electrons 
in the wind-backflow inside the radio cavity rapidly becomes
randomized after leaving the shock spot, 
possibly due to plasma microinstabilities.
Relatively few of the cosmic ray electrons in the backflow
actually pass though the bright spot; most escape directly
from the shock spot in other directions 
without experiencing a strong 
bright spot compression. 
}

Figure 10 shows the observed and computed radio emissivity 
$\epsilon_{\nu}$ at 1.345 GHz for the random field.
It is immediately seen that observed and computed emissivities  
agree quite well, differing by only 0.3 dex.
Precise agreement could 
be achieved by modifying some of our assumptions and 
approximations.
In addition, the slow downward trend of the maximum emissivity 
along the boundary backflow (decreasing $z$) is in 
excellent agreement with Cygnus A observations (Figure 6).

Table 2 lists the spatial coordinates of positions
($r,z$) of the 1.345 GHz emissivity peaks for each value of $z$.
Also listed are the field strength, emissivity
and backflow velocity, each evaluated at $r,z$.
The field strength is nearly uniform along the 
backflow and the computed values are almost identical 
to the lobe field $20\mu$G observed by Yaji et al. (2010).
Both these results 
provide excellent additional 
support for the random field model and indirect 
support for our assumptions of pitch angle scattering and 
viscosity.

Profiles of the CR energy density $e_c$ and 
(random field) magnetic energy 
density $u_B$ in the $r$-direction (at selected $z$) are shown in 
Figure 11.
The spatial uniformity of $e_c(r,z)$ across the lobe,  
required for pressure balance, by itself would suggest 
that the synchrotron emissivity from the Cygnus A lobes 
should be uniform, resulting in lobe limb-darkening 
not limb-brightening. 
However, 
synchrotron emission varies approximately as 
$e_cu_B \propto B^2$, 
therefore the peaks in $u_B$ seen in Figure 11
are critical to explain the observed limb-brightening.
Concentrations of magnetic energy density near the 
lobe boundaries, particularly in the region 
$z \approx 30 - 50$ kpc, 
occur in the rapidly moving boundary backflow 
of newly introduced electron CRs.

The declining magnetic field energy density at large $r$ in the 
$r$-profiles in Figure 11 can be easily understood 
as lobe confinement 
near the lobe-cluster gas boundary, 
but the sharp decline in $u_B$ at small $r$ 
inside the lobe depends critically on 
the evolutionary history of the Cygnus A shock spot. 
Figure 9 shows that the gas density in the shock spot 
$\langle \rho_{ss}(t)\rangle$ at early times can be 
several orders of magnitude 
larger than the typical lobe gas density computed 
at $10^7$ years, 
$\langle \rho_{lobe}(t)\rangle \sim 10^{-29}$ gm cm$^{-3}$.
Because of adiabatic expansion, 
the random magnetic field that entered the lobes from 
the shock spot at early times is lowered by a factor 
$(\langle \rho_{lobe}(t)\rangle / \langle \rho_{ss}(t)\rangle)^{2/3}$  
which is typically much less than unity. 
Since radio synchrotron emissivity varies as
$e_cu_B \propto B^2$, 
this explains (i) why 
the field in older regions of the radio lobe closer to the 
jet axis is so small and (ii) why synchrotron emission 
near the center of the radio lobes is so much less than that 
in the boundary backflow 
where the field reduction factor 
$\langle \rho_{lobe}(t)\rangle / \langle \rho_{ss}(t)\rangle$ 
is much closer to unity.
A sharply increasing magnetic field strength 
(and radio emission) with distance 
from the jet axis appears to be a genetic feature 
regardless of the magnetic field morphology. 
For example, this feature is seen when 
a toroidal magnetic field
is computed with the induction equation
(cf. Fig. 6 in MG12).
In both cases, magnetic field variations 
inside the radio lobes 
are essential in explaining radio lobe limb-brightening. 

Figure 12 illustrates the detailed energy and emission 
spectra of radio-synchrotron emitting electrons 
at five positions of maximum $\epsilon_{\nu}$ along 
the boundary backflow listed in Table 2.
The radio emissivity drops abruptly at $z \sim 34$ kpc 
due to radiative losses. 
Most of the 1.345 GHz emission from the boundary backflow 
in Figure 12 
behaves as expected, slowly decreasing with distance
from the shock spot.

Finally, in Figure 3 we see a small radio polarization 
in the boundary backflowing region 
which is often roughly parallel to the local cavity boundary. 
Assuming random field morphology, this can be understood 
as a competition between the instabilities that 
randomize the field and stretching of the random field 
as a result of the local transverse velocity gradient
perpendicular to the backflow. 
By this means, 
the shearing backflow is able to maintain a small apparent 
field along the flow direction, 
accounting for the backflow polarization seen in Figure 3.

\section{A second dynamical model matching the currently 
observed cavity volume}

As discussed in the Introduction, 
in choosing key parameters for 
the dynamical models of MG12 we did not attempt to 
slavishly duplicate all observed properties of Cygnus A.
In this sense our computed models are semi-quantitative.
Adopting an order-of-magnitude CR luminosity of 
$L_{cr} = 10^{46}$ erg s$^{-1}$, the MG12 calculations 
generated a 
(single) cavity volume at time 10 Myrs which is only 
about 55 percent of the observed volume  
$V_{obs} = 4.81 \times 10^4$ kpc$^3$,
determined from the measured size of 
a Cygnus A radio lobe assuming axisymmetry.
To test the results of our synchrotron emissivity 
results, we re-calculated a second dynamical model 
in which $L_{cr} = 2.65 \times 10^{46}$ erg s$^{-1}$
for which $V_{cav} \approx V_{obs}$ after 10 Myrs,
leaving all other initial parameters unchanged.

Figure 13 shows that this cavity-volume preserving 
model is able to fit the 1.345 GHz emissivity variations 
in the cavity just as well or better than the MG12 
model shown in Figure 10. 
Sub-kpc random magnetic fields are assumed in both figures.
Evidently our computed radio emissivities are rather 
insensitive to $L_{cr}$.
Another feature of this second dynamical model with 
enhanced relativistic energy density in the shock spot is that 
the offset separation between the shock spot and bright spot 
has increased to 2.5 kpc.

\section{Jet and cavity in Cygnus A are dominated by 
electron pairs}

The fraction of cosmic ray electron pairs that contribute to 
inflating the Cygnus A cavity can be found by comparing 
the energy density of {\it radiating} relativistic electrons 
in the current shock spot $e_{ce}$ (found from nearby 
bright spot observations) 
with the computed value of the {\it total} 
current cosmic ray 
energy density in the shock spot $e_c$
which includes both relativistic protons and electrons. 

The total energy density of {\it radiating} relativistic electrons
in the Cygnus A shock spot can be found from the current 
energy density observed by Stawarz et al. (2007) 
in the bright spot $e_{ce,bs} \propto K_{bs}$ 
for which $K_{bs} = 1.1 \times 10^{-4}$ cm$^{-3}$.
Our dynamical models (as in Fig. 5) can be used to estimate 
how the density normalization coefficient $K$ 
varies between shock and bright spots:
$K_{ss} = (\langle e_{c,ss} \rangle 
/ \langle e_{c,bs} \rangle) K_{bs}$
where volume-averaged values are determined for both 
shock spot grid zones and two similar 
grid zones at the bright spot.

The total energy density
of {\it radiating} relativistic electrons $e_{ce}$
in the Cygnus A shock spot is
\begin{equation}
e_{ce,ss} =
m_e c^2 \int_{\gamma_{cut}}^{\infty} \gamma n_{ss}(\gamma)d\gamma~~~
{\rm erg~cm}^{-3}
\end{equation}
where $n_{ss}(\gamma)$ is given by equation (23) 
and the local value of $K$ at the shock spot is $K_{ss}$.
Electron energy spectra in FRII sources typically have
low energy cutoffs
$\gamma_{cut} \sim 10^2 - 10^4$
(e.g. Mocz et al. 2011).
While no low energy cutoff has been observed in Cygnus A,
it might have escaped detection if $\gamma_{cut} \lta 100$.

Using the fiducial Cygnus A dynamical model from MG12
and a correction from observations of the bright spot of 
$K = K_{ss} = 1.38K_{bs}$,
the total integrated shock spot energy density in electrons is 
$e_{ce,ss} = 1.5 - 1.3 \times 10^{-8}$ 
erg cm$^{-3}$ for $1 < \gamma_{cut} < 100$. 
This range of electron energy densities 
is remarkably close to the 
current total relativistic energy density in the shock spot, 
$e_{c,ss} = 1.8 \times 10^{-8}$ erg cm$^{-3}$ from Figure 4
required to inflate the Cygnus A cavity in the MG12 dynamical 
model for which $L_{cr} = 10^{46}$ erg s$^{-1}$.

However, for a more accurate computation of $e_{c,ss}$
we can use the second dynamical model 
with $L_{cr} = 2.65 \times 10^{46}$ erg s$^{-1}$ that 
reproduces the observed volume of the radio lobe cavity 
at time 10 Myrs. 
For this model the computed zone-averaged total cosmic 
ray energy density in the shock spot is
$e_{ce,ss} = 2.2 - 2.5 \times 10^{-8}$ erg cm$^{-3}$
for the same range of $\gamma_{cut}$ 
and $K_{ss} = 2.31K_{bs}$. 
By comparison, the total cosmic ray energy density 
computed for this dynamical model at time 10 Myrs is 
$e_{c,ss} = 2.5 \times 10^{-8}$ erg cm$^{-3}$ from a 
figure analogous to Figure 4. 
For this solution the agreement is even closer, 
i.e. $e_{c,ss} \approx e_{ce,ss}$. 
Consequently, most or all of the relativistic particle energy 
inside the shock spot needed to inflate the cavity 
is contained in radiating electrons, 
i.e. the shock spot -- and by extension the jet and radio cavity -- 
are filled with a relativistic pair plasma.

This result is unaffected by the morphology of the field 
(toroidal or random) or by energy losses in the cavity due to
synchrotron losses. 
Most of the contribution to the integral for $e_{ce,ss}$ 
comes from $\gamma \sim \gamma_{cr} = 2000$
where the electron lifetime in the cavity
$t_*$ exceeds the age of Cygnus A for
$B = 20\mu$G, the cavity field estimated
by Yaji et al. (2010).
Consequently, synchrotron losses are not expected
to significantly degrade the cosmic ray energy density $e_c$
inside the radio cavity, and that was the assumption made
in the gasdynamical computations. 

Even a few relativistic protons in the shock spot 
having $\gamma \sim 2000$ 
would disrupt the near equality $e_{c,ss} \approx e_{ce,ss}$
found here.
Each relativistic proton has a contribution to the integral 
for $e_{cp,ss}$ that is 
$m_p/m_e \sim 1800$ times that of a relativistic electron 
with the same energy. 
Because of the low energy slope $p_1 = 1.5$ of $n(\gamma)$ 
in the shock spot, 
low energy relativistic electrons do not dominate 
the energy density $e_{ss}$.

\section{Foreflow synchrotron emission}

The heavy line contours 
in the upper panel of Figure 14 show an image of the 
projected radio flux  
at 1.345 GHz from Cygnus A at 10 Myrs, 
$\int \epsilon_{\nu}ds$, i.e. the computed emissivity 
integrated along the line of sight.
$L_{cr} = 10^{46}$ erg s$^{-1}$ is assumed.
The virtual jet moves 
along the $z$-axis from the left 
and shocks near $z = 60$ kpc. 
In Figure 14 we assume that the quasi-toroidal fields 
observed in the shock spot maintain the same morphology 
when they are compressed in the nearby 
bright spot; the enhanced field accounts for 
the strong radio flux from the bright spot.
Radiation from the bright spot comes from relativistic 
electrons with $\gamma > \gamma_{cr}$ having 
$n(\gamma) \propto \gamma^{-p_2}$.
Consequently, the bright spot radio emissivity increases 
rapidly with the field strength,
$\epsilon_{\nu} \propto B^{(1+p)/2} = B^{2.15}$ 
and $B$ increases rapidly toward the tip of the 
cavity (Fig. 5). 

The light line contours in 
the upper panel of Figure 14 show 
$\int e_c ds$ which is proportional to the (so far unobserved) 
surface brightness of 
IC-CMB X-ray emission from up-scattered CMB photons.
If the radio synchrotron image is simply approximated 
with $\int e_cu_Bds$, as in Figure 9 of MG12,
the radio emission from the offset bright spot 
in the forward direction 
is coextensive with $\int e_cds$.
However, in the upper panel of Figure 14 the radio synchrotron 
emission at 1.345 GHz drops off more rapidly 
than $\int e_cu_Bds$ toward the leading tip of the cavity.
The distance between peaks of 1.345 GHz and $\int e_cds$ 
emission is only about 1 kpc. 

A close examination reveals that this truncation of 
the 1.345 GHz image near the tip of the cavity 
is due to radiative losses,
causing $n(\gamma) \rightarrow 0$ in the small tip region 
where $\int e_cu_Bds$ contours 
extend further in the $z$-direction than those for $\int e_cds$.
In principle, the 
arc-shaped radio synchrotron emission from the bright 
spot may not always precisely 
define the cavity-cluster gas interface 
at all radio frequencies.
Of course the details of this transition depend on 
assumptions made in Section 3. 
In any case, the total energy density 
of synchrotron-emitting electrons observed 
by Stawarz et al. (2007) in the 
bright spot may be slightly underestimated because 
of radiative losses.

\section{Cygnus A at a later time}

The proximity of shock and bright spots in the upper panel of 
Figure 14 definitely does not indicate that the 
distinction we make between these two regions is 
unnecessary or unphysical.
To illustrate this, in 
the lower panel of Figure 14 we show the projected 
appearance of Cygnus A at 20 Myrs, twice its current 
age, assuming that the shock spot maintains its 
current properties (CR luminosity, velocity, magnetic field,
etc.) in the future.
Because of the negative gas density gradient in the Cygnus A 
cluster, the shock and bright spots are now offset by 
an easily observable 3 kpc. 
The bright spot radio display is also larger 
since the foreflowing shock spot wind evacuates a larger region  
as the density of the local cluster gas decreases. 
This is consistent with the observation that the size 
of FRII bright spots systematical increases with projected core-hotspot 
distance (Hardcastle, Alexander, Pooley \& Riley 1998).
Notice that at age 20 Myrs a small amount of radio emission 
at 1.345 GHz appears at the shock spot.

For an even more extreme example Figure 15  
shows 1.4 GHz radio synchrotron (contours) 
and non-thermal X-ray (gray scale) emission 
observed in the FRII source 4C74.26 
by Erlund et al. (2010). 
X-rays are observed from both the bright spot (SSC) and shock 
spot (X-ray synchrotron?) but the bright spot clearly dominates 
at radio frequencies.
Both spots are elongated.
In 4C74.26 the bright radio spot is 
in very low density gas 500 kpc from the center 
of its cluster and the physical separation between the 
shock and bright spots is 19 kpc.
The dashed line shows one possibility for the jet trajectory 
needed to activate the observed shock spot, the dotted line 
shows another,
The change in orientation between the X-ray 
shock spot (roughly perpendicular 
to the jet) and the X-ray bright spot indicates that the bright 
spot is formed as the jet strikes the  
radio cavity-cluster gas interface that is inclined to the 
jet direction.

\section{Conclusions}

Observations of the Cygnus A radio cavity indicate 
that the magnetic field is nowhere strong enough to 
directly influence the gas plus CR dynamics, 
$u_B < e_c + e$.
Nevertheless, the frozen-in field connects the dynamical 
motion of relativistic and non-relativistic components 
in our two-fluid hydrodynamical models of 
Cygnus A evolution (MG12).
Because the magnetic field evolves passively with the 
post-shock hydrodynamic flow, a variety of different 
field evolution models can be investigated and 
compared with observation without re-computing the gasdynamics.
The goal is to find transport properties, 
field morphologies, etc. that best fit detailed observations.
By this means we hope to understand better 
the complex, poorly understood small scale
physics in weakly magnetized plasmas dominated 
by relativistic energy density. 

\vskip.1in\noindent$\bullet$
A new terminology is proposed that avoids use of the 
standard ``hotspot'' designation near the tips of 
FRII jets.
The ``shock spot'' is the immediate post-shock region that 
serves as the source of a wind that transports 
cosmic rays, gas and magnetic 
field into the Cygnus A radio cavity. 
The "bright spot'' is the much more radio luminous region where
the shock spot wind impacts against the dense (shocked) 
cluster gas just ahead. 

\vskip.1in\noindent$\bullet$
The Kelvin-Helmholtz instability 
(KHI) in FRII radio lobes must be damped to match the 
smooth appearance of observed radio cavities and the 
monotonic variation of observed synchrotron spectral 
ages inside the cavity.
It is unlikely that the KHI in Cygnus A can be stabilized 
by large-scale magnetic fields along the direction of the boundary 
backflow inside the radio cavity.
For stability the Alfv\'en speed $v_A = B/(4 \pi \rho)^{1/2}$
must exceed the velocity difference across the shearing 
layer. 
To achieve this in Cygnus A either (i) the magnetic field in this 
one direction must be $\sim 10$ times larger than 
the total field observed 
in the radio cavity or (ii) the density $\rho$ of non-relativistic 
gas must be 100 times smaller than assumed in the MG12 
calculation. 
But the ultra-high backflow velocity at these ultra-low densities may  
cause the radio synchrotron ages in the cavity backflow 
to be smaller than observed.

\vskip.1in\noindent$\bullet$
If gas inside the radio lobe has a  small viscosity, 
as we assume,  
the KHI can be damped without strong magnetic fields.

\vskip.1in\noindent$\bullet$
Radio lobe limb-brightening is due largely to the increased 
lobe magnetic field in the boundary backflow region.
The wind that emanates from the shock spot carries 
non-relativistic gas, magnetic field and  
cosmic ray electrons, all of which flow back in the anti-jet 
direction along the radio lobe boundaries. 
The radio synchrotron emissivity depends on the product 
of the cosmic ray and magnetic energy densities, 
$e_cu_B$.
However, if $e_c$ 
dominates the pressure in the radio lobes, as we assume,
$e_c$ must be nearly uniform across 
(and throughout) the radio lobes to balance the pressure 
in the local external (shocked) cluster gas. 
Consequently, radio limb-brightening is not due to 
spatial variations in the CR electron density but instead 
to the varying magnetic field which is larger 
near the lobe boundaries.

Assuming a random sub-grid field morphology,
the field anywhere in the lobe has adiabatically 
expanded from the value it had 
when it left the shock spot.
But the field is frozen into the non-relativistic 
gas as it expands from the 
shock spot to a lower density in the observed lobe 
following $B \propto \rho^{2/3}$. 
The current gas density in the lobe that arises 
in the shock spot wind 
$\rho_{lobe}(t)$ is approximately 
uniform inside the lobe.
By assumption, the field in the shock spot remains 
constant with time, but
the shock spot gas density $\rho_{ss}$ decreases dramatically 
as the lobe evolves into the cluster gas.
Because of this, the field in the lobe varies as 
$(\rho_{ss}(t))^{-2/3}$ 
which increases with time during the lobe evolution 
and decreases with the age of the CRs.
Therefore, the lobe field that emerged from the  
shock spot most recently 
(located in the boundary backflow)
is larger than fields that entered the lobe 
at previous times. 
This enhances the field and synchrotron emissivity 
in the boundary backflow and 
explains the observed radio lobe limb-brightening.
Evidently a similar argument also explains large-scale
limb-enhanced magnetic fields and limb-brightening 
in lobes computed using the induction equation.

\vskip.1in\noindent$\bullet$
It is shown how the analytic expression for the evolution 
of radio synchrotron electrons due to uniform expansion and 
radiation losses can be adapted to estimate the electron 
energy distribution $n(\gamma,{\bf r},t)$ and specific 
radio synchrotron emissivity $\epsilon_{\nu}({\bf r},t)$ 
at every 
grid zone and time inside the radio cavity.
These local functions are quite well determined from 
only three computed 
hydrodynamic arrays: gas density $\rho$, magnetic field $B$
and the advected time since local synchrotron-emitting 
electrons left the post-shock region. 

\vskip.1in\noindent$\bullet$
For the first time the local energy spectrum $n(\gamma)$
and the emissivity $\epsilon_{\nu}$ 
of radio-synchrotron electrons
can be quantitatively computed 
throughout the radio cavity only from observations 
of the bright spot (when corrected to shock spot values) 
and simple assumptions about shock spot 
evolution over time.

\vskip.1in\noindent$\bullet$
Parameters describing 
the double power law $n(\gamma)$ observed in the Cygnus A 
bright spots need to be corrected to conditions in the nearby  
shock spot region. 
The shock spot is the source of 
electrons in the bright spot and the wind that 
inflates the radio cavity. 
This correction can be made 
from gasdynamical calculations. 

\vskip.1in\noindent$\bullet$
When this correction is made, the total energy density of 
relativistic synchrotron electrons in the current shock spot 
$e_{ce,ss}$ matches the total relativistic 
energy density $e_{c,ss}$ required to inflate 
the Cygnus A cavity to its current volume. 
This indicates that the relativistic fluid in radio 
cavities consists of electron pairs rather than  
a relativistic electron-proton plasma. 
A pure relativistic pair plasma 
(possibly with some entrained non-relativistic gas 
acquired during jet-gas interactions with the thermal filament) 
is physically attractive since it may be consistent with 
a purely electromagnetic origin 
of FRII jets (e.g. Blandford 2008). 

\vskip.1in\noindent$\bullet$
The computed radio synchrotron emission from 
Cygnus A cavities having purely toroidal magnetic 
fields is a very poor fit to VLA observations.
Boundary backflows decelerate 
due to small negative radial pressure gradients 
inside the lobes. 
During deceleration, 
the toroidal field grows exponentially, causing 
the synchrotron emissivity to increase unrealistically 
along the backflow.
Toroidal fields rapidly grow to exceed the fields 
observed in the Cygnus A radio cavity.
Evidently the radial component of the field $B_r$, 
which is also transverse to the decelerating boundary backflow,
would also increase in the same undesirable  
way along the backflow.
This unfavorable outcome indicates that the 
usual MHD codes that solve the induction equation 
for the magnetic field on grid scales or larger 
are unlikely to 
provide realistic solutions inside radio cavities.

\vskip.1in\noindent$\bullet$
Consequently, we consider a field that is random 
on subgrid scales and which advects as 
$B_{ran} \propto \rho^{2/3}$.
Such a field may result from small scale 
turbulence due to plasma instabilities.
When saturated, the magnetic energy 
density $u_B = B_{ran}^2/4\pi$
may approach a low value comparable to the 
non-relativistic energy density $e$ which is generally 
less than the relativistic energy density $e_c$.
Random subgrid activity would also be consistent 
with the viscous transport of non-relativistic gas, 
a likely source of viscous KH damping.
Using a simple model for $B_{ran}$, 
we find that the radio synchrotron emissivity 
inside the Cygnus A cavity can be fit quite well.
$B_{ran}$ is almost constant along the boundary 
backflow and has a value $\sim 20\mu$G that is 
almost identical to values observed in the radio 
cavity.
The radio emissivity predicted by the random 
field model slowly decreases along the backflow, 
similar to VLA observations, with emissivities that 
quantitatively match these same observations.

\vskip.1in\noindent$\bullet$
If the jet and shock spot parameters remain fixed, 
the separation between the shock spot and bright spot 
increases with time, becoming about three times larger 
when Cygnus A is at twice its current age.
This can be explained by the reduced cluster gas density 
at larger distances from the cluster core, 
allowing the shock spot wind to evacuate a larger 
local volume. 

\vskip.1in
\acknowledgements
Studies of the evolution of Cygnus A are supported 
by Chandra theory grant TM1-12009X and by 
NSF grant AST-0807724 
for which we are very grateful.
David A. Clarke and Martin Hardcastle provided helpful 
thoughts and advice.
We extend special thanks 
to Sandra Holeman for recovering our grant funding.

\clearpage

\begin{deluxetable}{cccrccc}
\tabletypesize{\scriptsize}
\tablecolumns{7}
\tablewidth{6.cm}
\tablecaption{BACKFLOW VARIATIONS FOR TOROIDAL FIELD
FOR VARIOUS $z$ AT $r$ OF MAXIMUM 1.345 GHz EMISSIVITY}
\tablehead{
\colhead{$k$} &
\colhead{$z$} &
\colhead{$r$} &
\colhead{$B_t$} &
\colhead{$(\epsilon_{1.345})_{max}$~ $\times 10^{34}$} &
\colhead{$(u_z/c)$} &
\colhead{$(u_r/c)$}  \cr
\colhead{} &
\colhead{(kpc)} &
\colhead{(kpc)} &
\colhead{($\mu$G)} &
\colhead{(erg/cm$^{3}$Hz)} &
\colhead{} &
\colhead{} \cr
}
\startdata
0 & 60 & 0.5 & 62  & 58 & -0.23  & 0.25 \cr
1 & 50 & 3.7 & 77  & 13 & -0.18  & 0.041 \cr
2 & 40 & 6.2 & 126 & 31 & -0.056 & 0.018  \cr
3 & 30 & 8.7 & 167 & 44 & -0.025 & 0.0087 \cr
4 & 20 & 8.7 & 151 & 22 & -0.0082  & 0.0013 \cr
\enddata
\end{deluxetable}


\clearpage

\begin{deluxetable}{cccrccc}
\tabletypesize{\scriptsize}
\tablecolumns{7}
\tablewidth{6.cm}
\tablecaption{BACKFLOW VARIATIONS FOR RANDOM FIELD
FOR VARIOUS $z$ AT $r$ OF MAXIMUM 1.345 GHz EMISSIVITY}
\tablehead{
\colhead{$k$} &
\colhead{$z$} &
\colhead{$r$} &
\colhead{$B_{ran}$} &
\colhead{$(\epsilon_{1.345})_{max}$~ $\times 10^{34}$} &
\colhead{$(u_z/c)$} &
\colhead{$(u_r/c)$}  \cr
\colhead{} &
\colhead{(kpc)} &
\colhead{(kpc)} &
\colhead{($\mu$G)} &
\colhead{(erg/cm$^{3}$Hz)} &
\colhead{} &
\colhead{} \cr
}
\startdata
0 & 60 & 0.5 & 62  & 181  &  0.14  & 0.27 \cr
1 & 50 & 3.7 & 23  & 0.96 & -0.18  & 0.041  \cr
2 & 40 & 6.2 & 23 & 0.82  & -0.051 & 0.015 \cr
3 & 30 & 8.7 & 22 & 0.61  & -0.020 & 0.0068 \cr
4 & 20 & 8.7 & 22 & 0.45  & -0.0082 & 0.0013 \cr
\enddata
\end{deluxetable}


\clearpage

\begin{figure} 
\centering
\includegraphics[width=4.4in,scale=0.45,angle=0]
{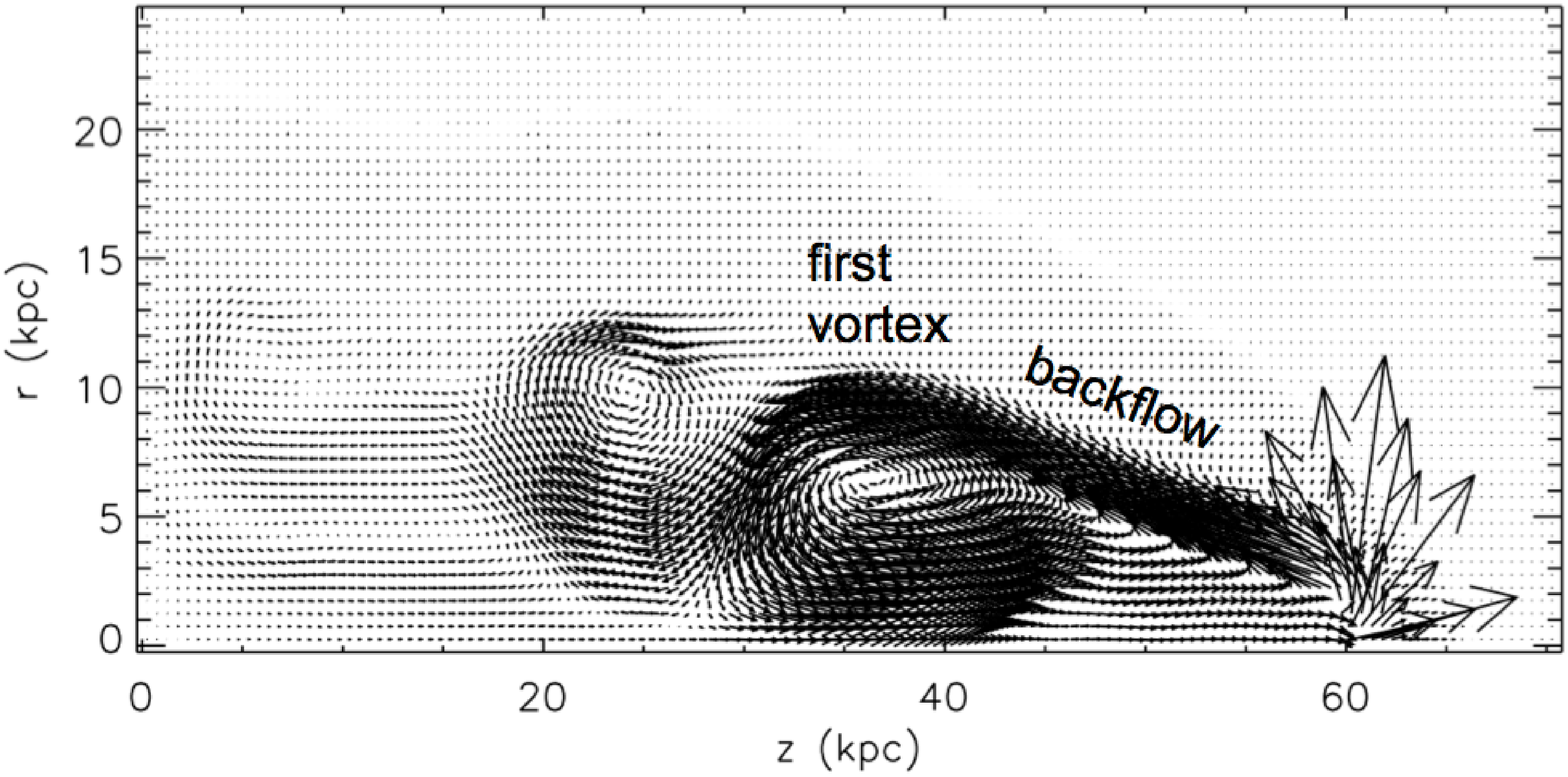}
\vskip-.3in
\vskip.05in
\includegraphics[width=4.15in,scale=.45,angle=0]
{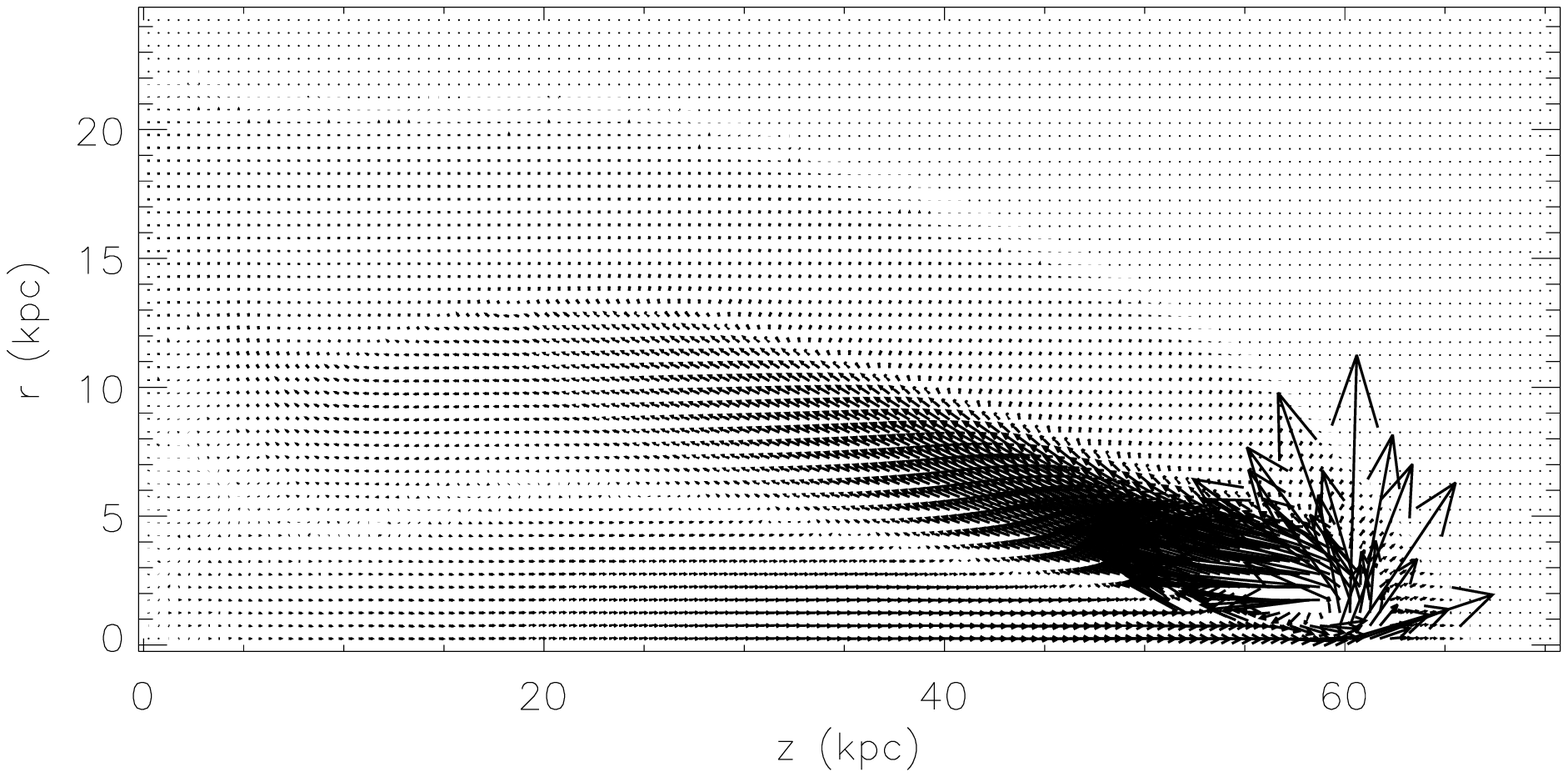}
\caption{
Two calculations of overlapping flow velocity vectors 
for the flow in Cygnus A after 10 Myrs. 
Both gasdynamical calculations, taken from MG12, 
are identical but without viscosity ({\it top panel}) 
and with a small viscosity,  
$\mu = 30$ gm cm$^{-1}$ s$^{-1}$ ({\it bottom panel}).
The first vortex in the top panel shows that the initial 
onset of KHI deviates the flow toward the symmetry 
axis of the cavity. 
Since this is the same sense as the local velocity 
shear at the interior surface of the backflow, 
the KHI first appears at 
the {\it internal} boundary of the backflow, not 
because of shear with the cluster gas.
}
\end{figure}

\clearpage

\begin{figure} 
\centering
\includegraphics[width=4.0in,scale=0.45,angle=0]
{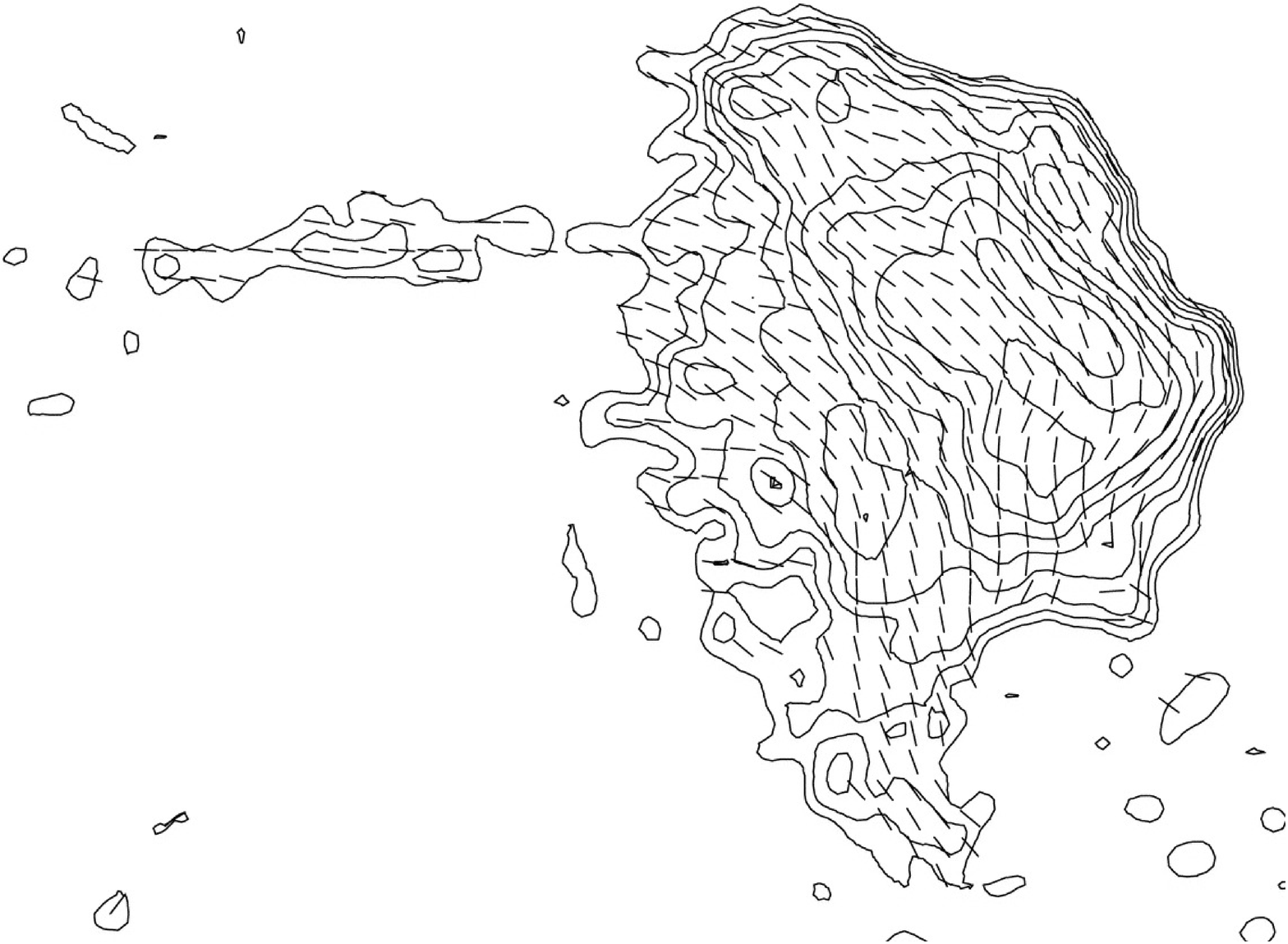}
\vskip.05in
\caption{
Intensity contours at $\nu = 43$ GHz
with polarization vectors ($\parallel$ 
to $B$) of the northwestern bright spot A 
from Carilli et al. (1999). 
The invisible jet is incident from the lower left.
The entire image is 
about $5.5 \times 7.4$ kpc
}
\end{figure}

\clearpage

\begin{figure} 
\centering
\includegraphics[width=4.0in,scale=0.45,angle=90]
{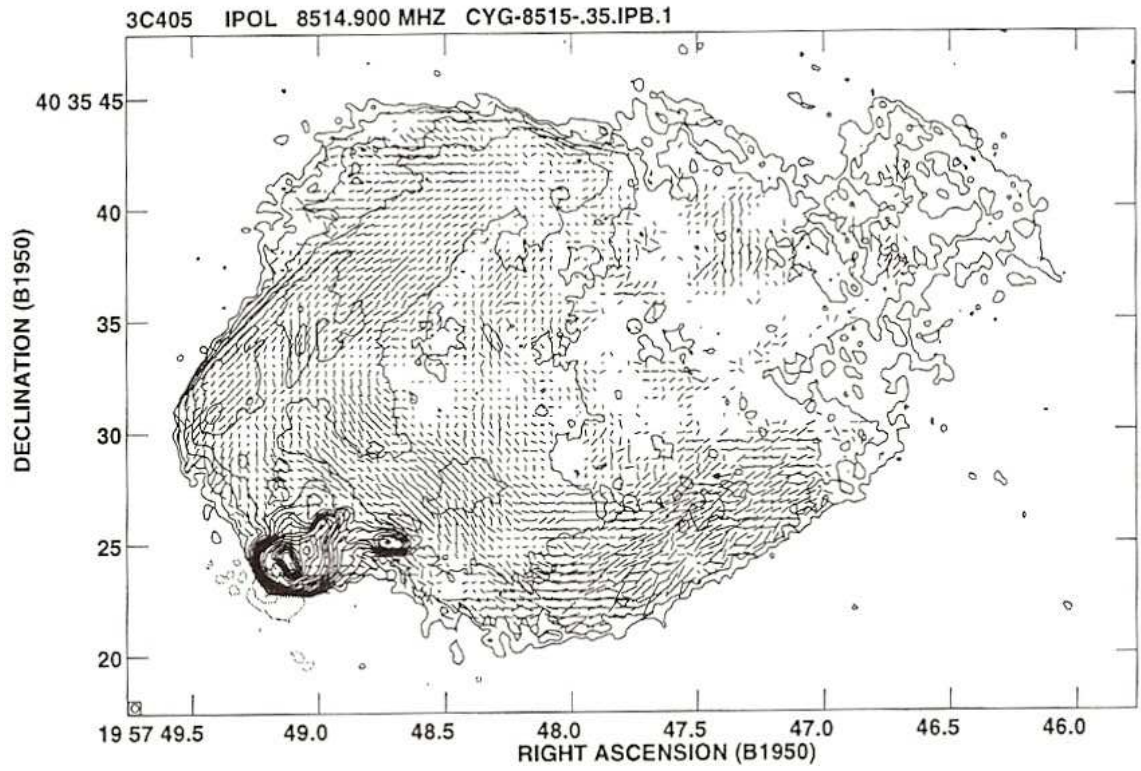}
\vskip.05in
\caption{
Intensity contours at $\nu = 8.5$ GHz
with polarization vectors ($\parallel$
to $B$) of the eastern radio lobe in Cygnus A
(Carilli \& Harris 1996).
The entire image is
about $30 \times 53$ kpc
}
\end{figure}

\clearpage

\begin{figure} 
\centering
\includegraphics[width=4.0in,scale=0.45,angle=270]
{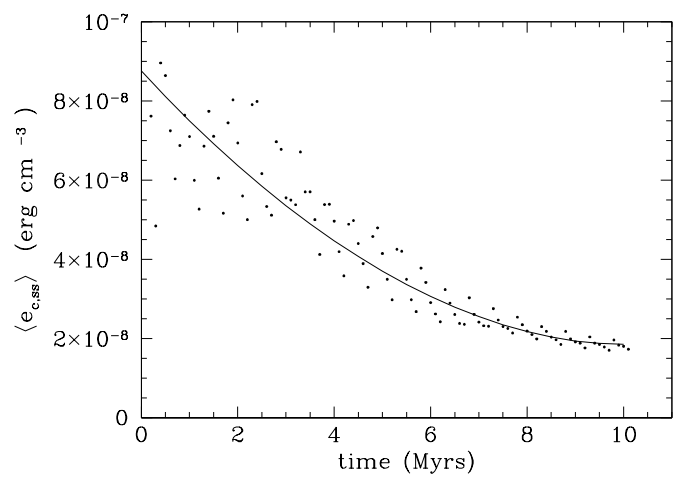}
\vskip.05in
\caption{
Time variation of the cosmic ray energy density in the 
shock spot with an analytic fit.
}
\end{figure}

\clearpage

\begin{figure}
:
\centering
\includegraphics[width=4.0in,scale=0.45,angle=0]
{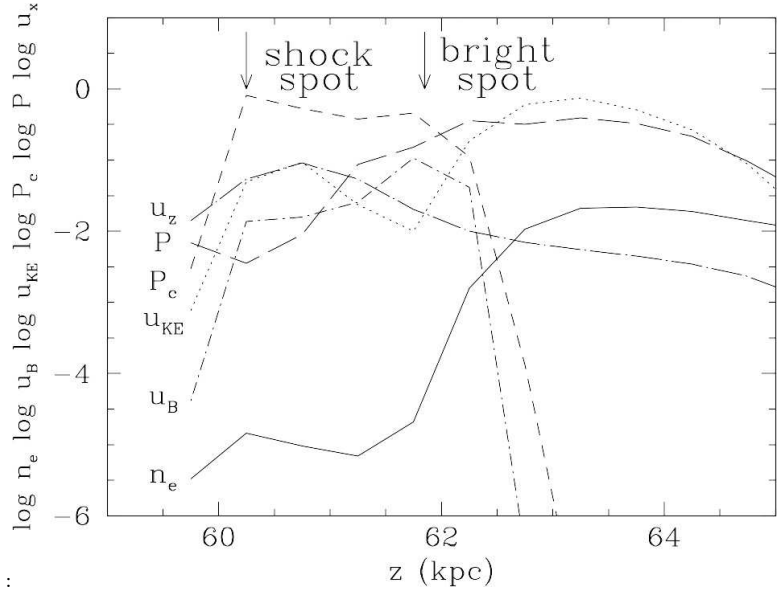}
\vskip.05in
\caption{
Detailed zone-by-zone profiles of the shock-bright spot structure
along the jet (symmetry) axis as in Figure 10 of MG12.
Values plotted are those of computational zones closest 
to the $z$-axis. 
The gas pressure $P = 2/e/3$, CR pressure $P_c = e_c/3$,
magnetic energy density $u_B = B^2/8\pi$,
and kinetic energy density $u_{KE} = \rho {(u_z)^2}/2$
are in cgs units increased by a factor of $10^8$.
The gas velocity $u_z(z)$ in cgs units
is reduced by a factor $10^{-11}$.
The CR energy density $e_c$ peaks at the shock spot
where energy is injected, but radio synchrotron emission
$\propto e_cu_B$ peaks at the bright spot, about 1.5 kpc
ahead.
}
\end{figure}

\clearpage

\begin{figure}
\centering
\includegraphics[width=4.0in,scale=0.45,angle=0]
{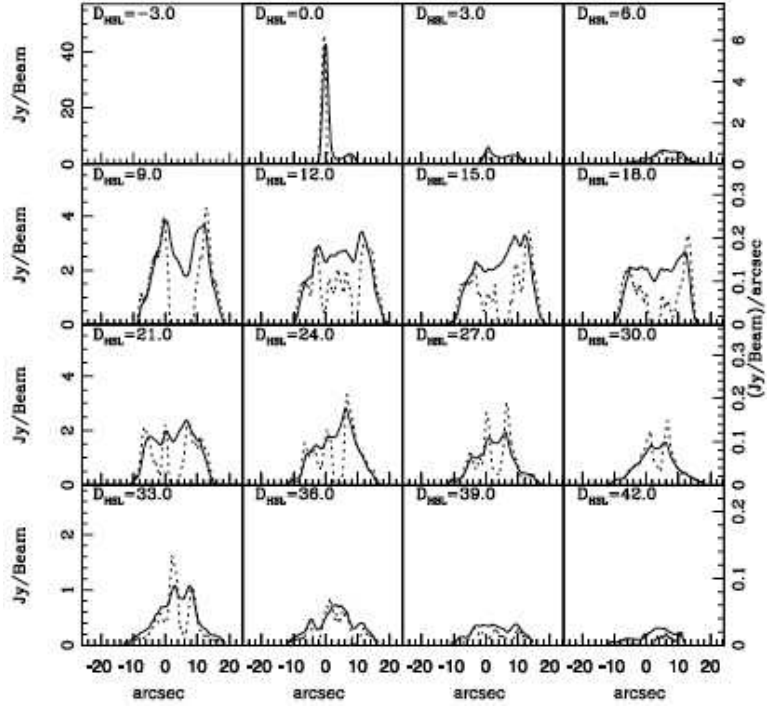}
\vskip.05in
\caption{
Solid lines are (complete) profiles of the 
observed radio synchrotron flux 
(in Jy/beam) at $\nu = 1.345$ GHz perpendicular to the mean jet direction 
in Cygnus A at 16 different distances from the eastern bright spot, 
$D_{hs} = 60 - z_{bs}$ kpc (Carvahlo et al. 2005).
Dashed lines show profiles of the corresponding 
radio emissivity given in units of $9.0 \times 10^{-34}$ 
erg cm$^{-3}$ Hz$^{-1}$ on the right. 
Note that the vertical scale changes below the first row.
The horizontal axis is shown in arcseconds or kpc 
(1\arcsec $\approx 1$ kpc).
}
\end{figure}

\clearpage

\begin{figure}
\centering
\includegraphics[width=4.0in,scale=0.45,angle=0]
{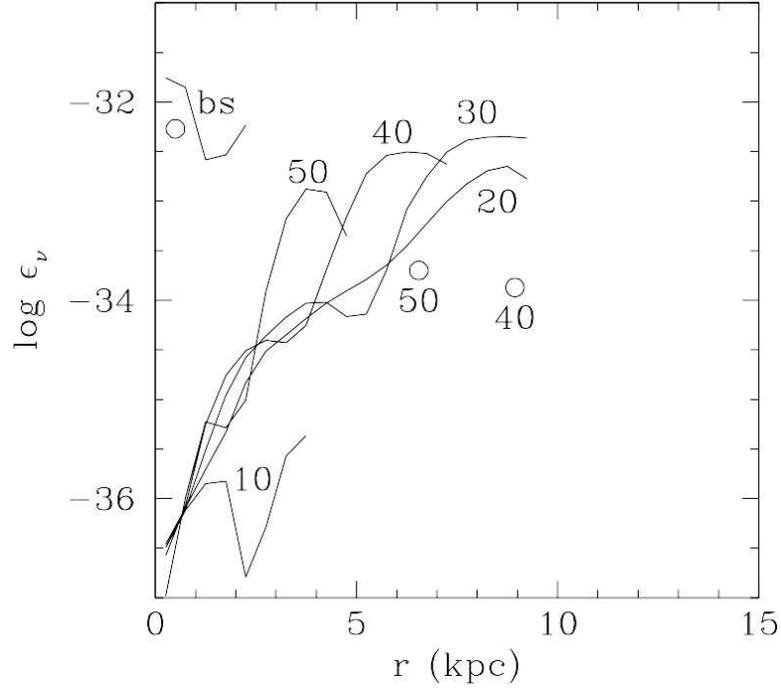}
\vskip.05in
\caption{
Transverse radio synchrotron emission profiles
for toroidal magnetic field.
Solid curves show the variation of computed emissivity  
$\epsilon_{\nu}(r)$ (erg cm$^{-3}$ Hz$^{-1}$) 
at 1.345 GHz 
perpendicular to the Cygnus A symmetry axis. 
Each curve is labeled with its distance $z$ in kpc 
from the cluster center or with ``bs'' the bright spot 
emissivity at $z_{bs} = 60$ kpc.
The computed 
radio emissivity peaks in the boundary backflow 
region for $50 \gtrsim z \gtrsim 30$ kpc, then drops sharply 
at $z = 20$ and 10 kpc.
The three open circles are the observed peak emissivities 
$\epsilon_{\nu}$ copied from Figure 6 and labeled with 
``bs'' or $z$ in kpc.
}
\end{figure}

\clearpage

\begin{figure}
\centering
\includegraphics[width=4.0in,scale=0.45,angle=0]
{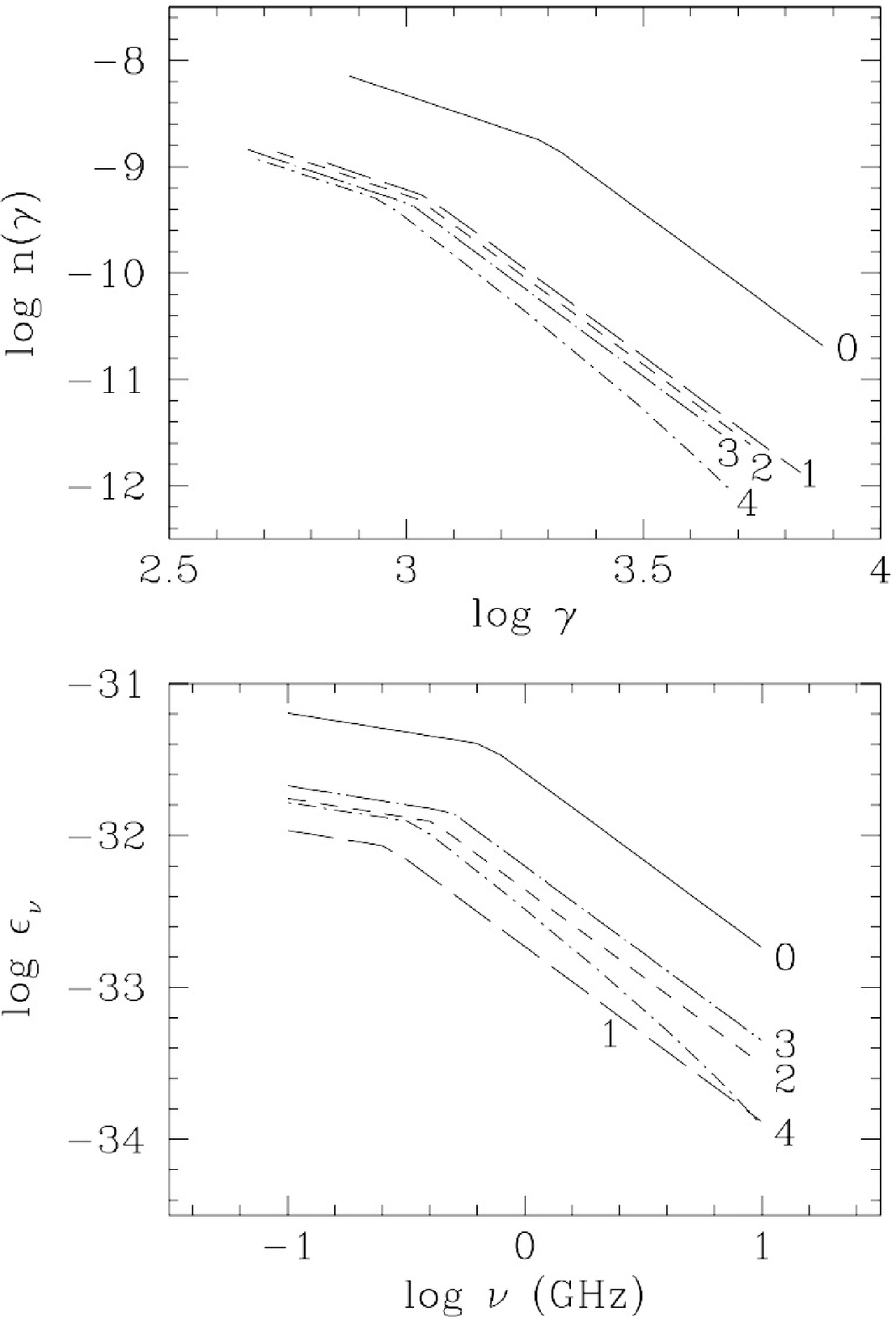}
\vskip.05in
\caption{
Energy distribution and emissivity for radio-synchrotron electrons 
in a toroidal field.
{\it Upper panel:} Computed electron number density $n(\gamma,r,z)$ 
cm$^{-3}$
in the shock spot (labeled ``0'') and at four $(r,z)$ positions 
of maximum $\epsilon_{\nu}$ along the backflow  
listed in Table 1.
{\it Lower panel:} Emissivity spectra 
$\epsilon_{\nu}(r,z)$ (erg cm$^{-3}$ Hz$^{-1}$)
computed at each of the five positions listed in 
Table 1.
}
\end{figure}

\clearpage

\begin{figure}
\centering
\includegraphics[width=4.0in,scale=0.45,angle=0]
{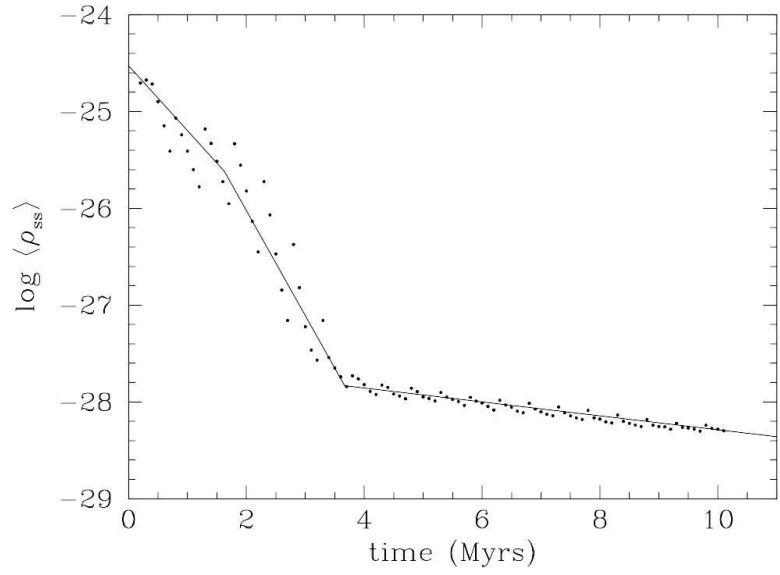}
\vskip.05in
\caption{
Variation of the gas density in the shock spot 
(gm cm$^{-3}$) with time during the Cygnus A evolution. 
The computational points are fit with three straight lines.
}
\end{figure}

\clearpage

\begin{figure}
\centering
\includegraphics[width=4.0in,scale=0.45,angle=0]
{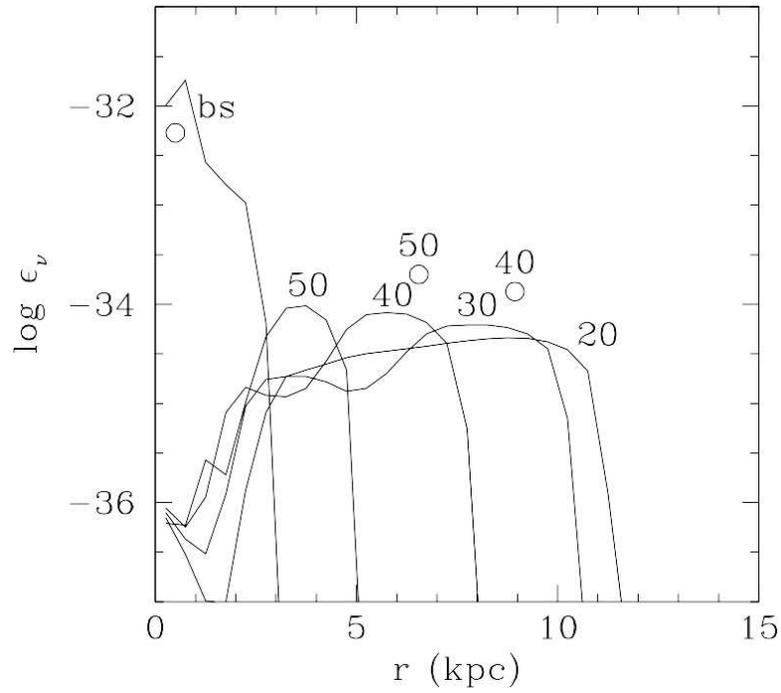}
\vskip.05in
\caption{
Radio synchrotron emission for a random small-scale magnetic field.
Solid curves show computed emissivity profiles
$\epsilon_{\nu}(r)$ (erg cm$^{-3}$ Hz$^{-1}$)
at 1.345 GHz
perpendicular to the Cygnus A symmetry axis.
Each curve is labeled with its distance in kpc
from the cluster center or with ``bs'' for the bright spot
emissivity at $z_{bs} = 60$ kpc.
The computed
peak emissivity slowly decreases along the boundary backflow
when $50 \gtrsim z \gtrsim 30$ kpc, then drops sharply
near $z = 20$ kpc.
The three open circles are the observed peak emissivities
$\epsilon_{\nu}$ copied from Figure 6, each labeled with
``bs'' or $z$ in kpc.
}
\end{figure}

\clearpage

\begin{figure}
\centering
\includegraphics[width=4.0in,scale=0.45,angle=0]
{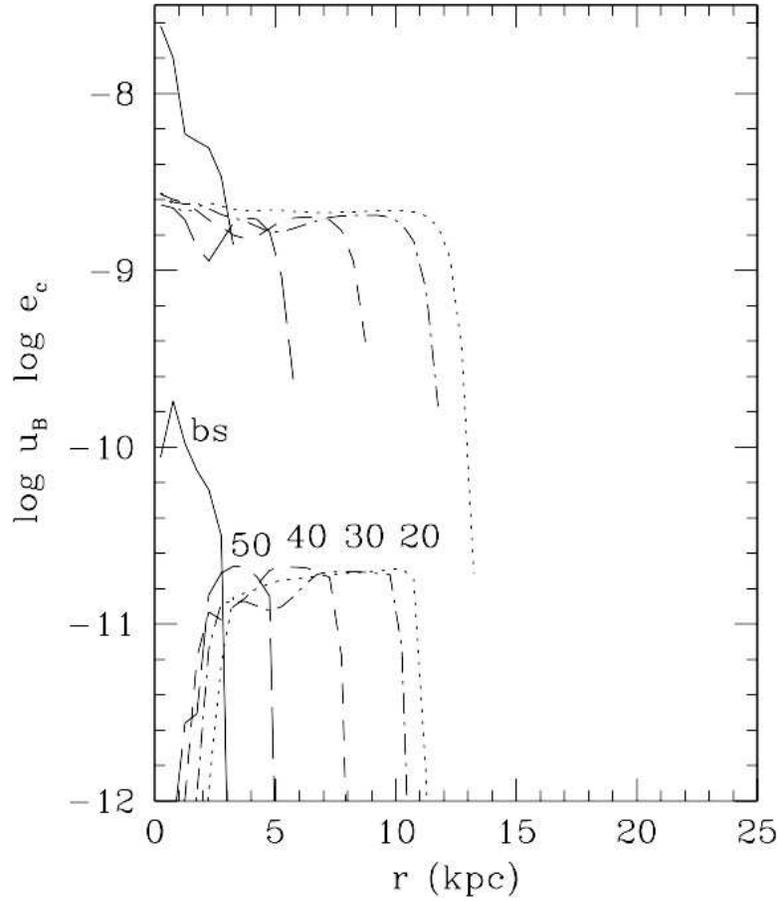}
\vskip.05in
\caption{
Transverse $r$-contours of CR energy density
$e_c(z,r)$ ({\it upper plots})
and random small-scale
magnetic energy density $u_B(z,r)$ ({\it lower plots})
in erg cm$^{-3}$ at time $10^7$ years.
From left to right each set of plots shows
$r$-profiles of $e_c(z,r)$ and $u_B(z,r)$
at $z = 20$, 30, 40, 50 and 60 kpc
(distance from Cygnus A core) shown respectively with
dotted, dash-dotted, short dashed, long dashed, and solid lines.
The strong central magnetic peak at $z = 60$ kpc
corresponds to the bright spot.
The $e_c(z,r)$ profiles remain approximately uniform across the
cavity to maintain approximate pressure equilibrium with the
(shocked) cluster gas.
By contrast, $r$-profiles of the magnetic energy density $e_c$
(each identified with its $z$ in kpc) peak near the lobe
boundary when $z \gtrsim 30$ kpc and become more uniform at
smaller $z$ near the cluster core.
The increase in the magnetic field near the lobe boundary
strongly enhances synchrotron emission from the boundary
backflow.}
\end{figure}

\clearpage

\begin{figure}
\centering
\includegraphics[width=4.0in,scale=0.45,angle=270]
{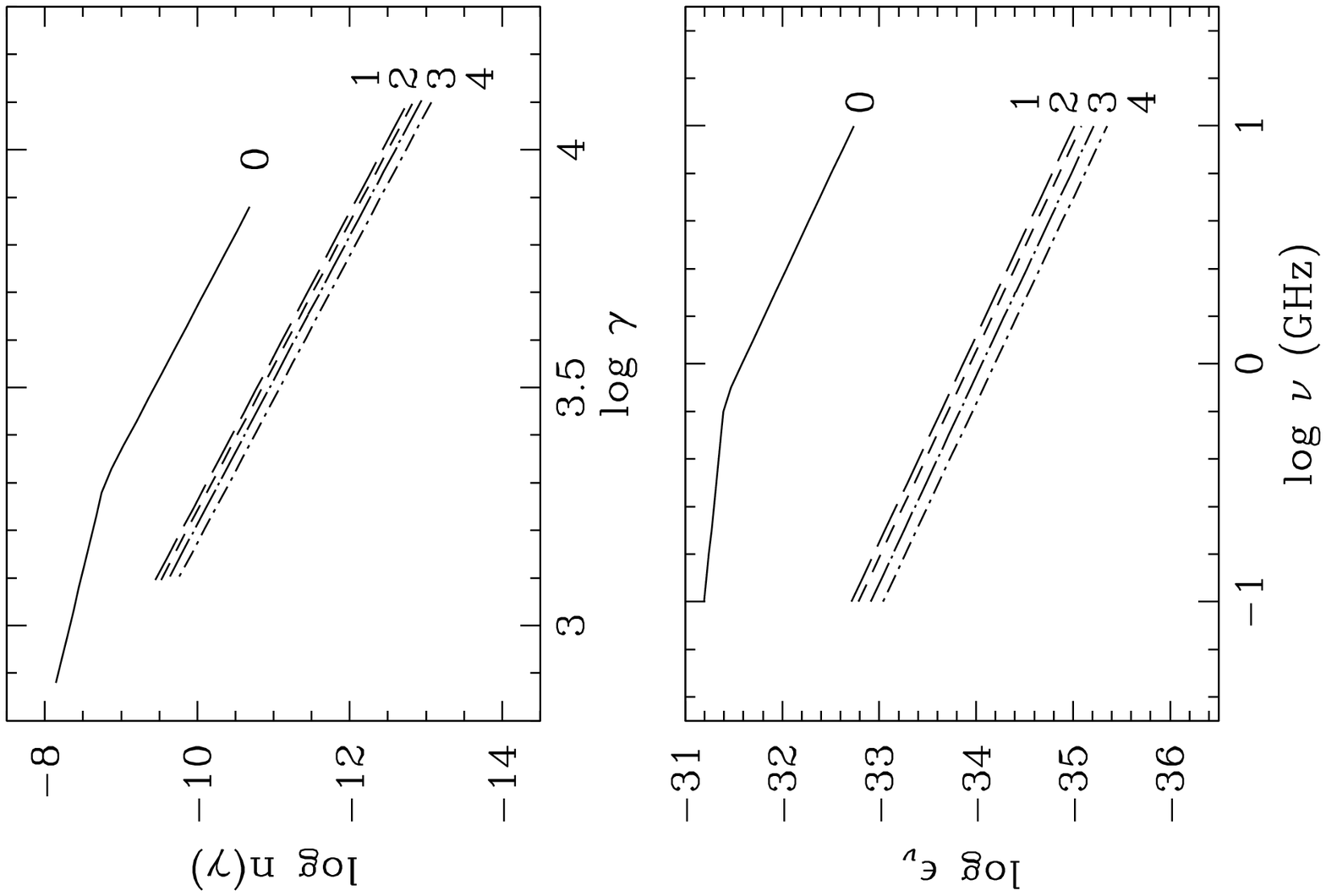}
\vskip.05in
\caption{
Energy distribution and emissivity spectra 
for radio-synchrotron electrons
in a random field.
{\it Upper panel:} Computed electron number density $n(\gamma,r,z)$
cm$^{-3}$
in the shock spot (labeled ``0'') and at the four $(r,z)$ positions
of maximum $\epsilon_{\nu}$ along the backflow
listed in Table 2.
{\it Lower panel:} Emissivity $\epsilon_{\nu}(r,z)$ (erg cm$^{-3}$
Hz$^{-1}$)
computed at each of the five positions listed in
Table 2.
}
\end{figure}

\clearpage

\begin{figure}
\centering
\includegraphics[width=4.0in,scale=0.45,angle=0]
{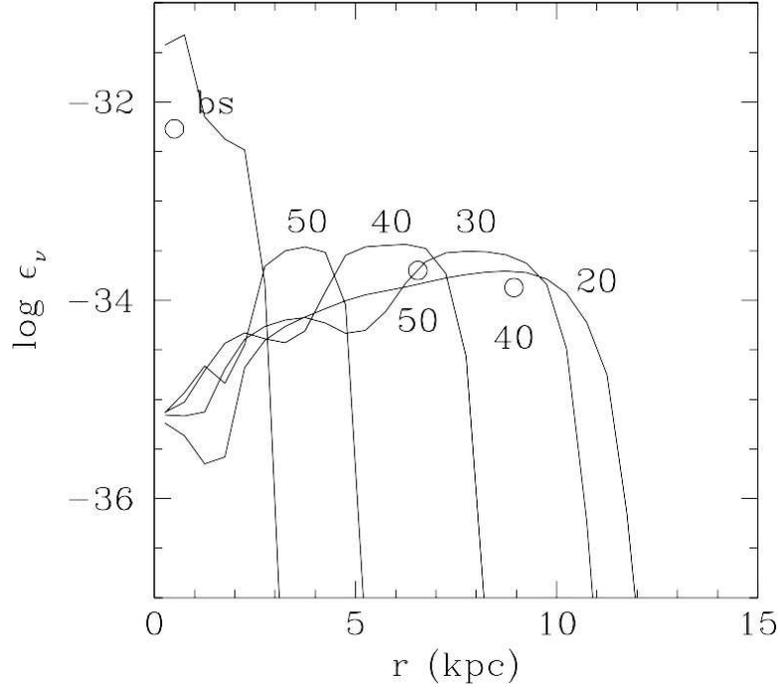}
\vskip.05in
\caption{
Radio synchrotron emission for a random small-scale magnetic field
for a second dynamical model with $L_{cr} = 2.65 \times 10^{46}$ 
ergs s$^{-1}$ and 
a radio cavity volume that 
matches that observed today at assumed age 10 Mys.
Solid curves show computed emissivity profiles
$\epsilon_{\nu}(r)$ (erg cm$^{-3}$ Hz$^{-1}$)
at 1.345 GHz
perpendicular to the Cygnus A symmetry axis.
Each curve is labeled with its distance in kpc
from the cluster center or with ``bs'' for the bright spot
emissivity at $z_{bs} = 60$ kpc.
The computed
peak emissivity slowly decreases along the boundary backflow
when $50 \gtrsim z \gtrsim 30$ kpc, then drops sharply
due to radiation losses near $z = 20$ kpc.
The three open circles are the observed peak emissivities
$\epsilon_{\nu}$ copied from Figure 6 and labeled with
``bs'' or $z$ in kpc.
}
\end{figure}

\clearpage

\begin{figure}
\centering
\includegraphics[width=4.0in,scale=0.45,angle=0]
{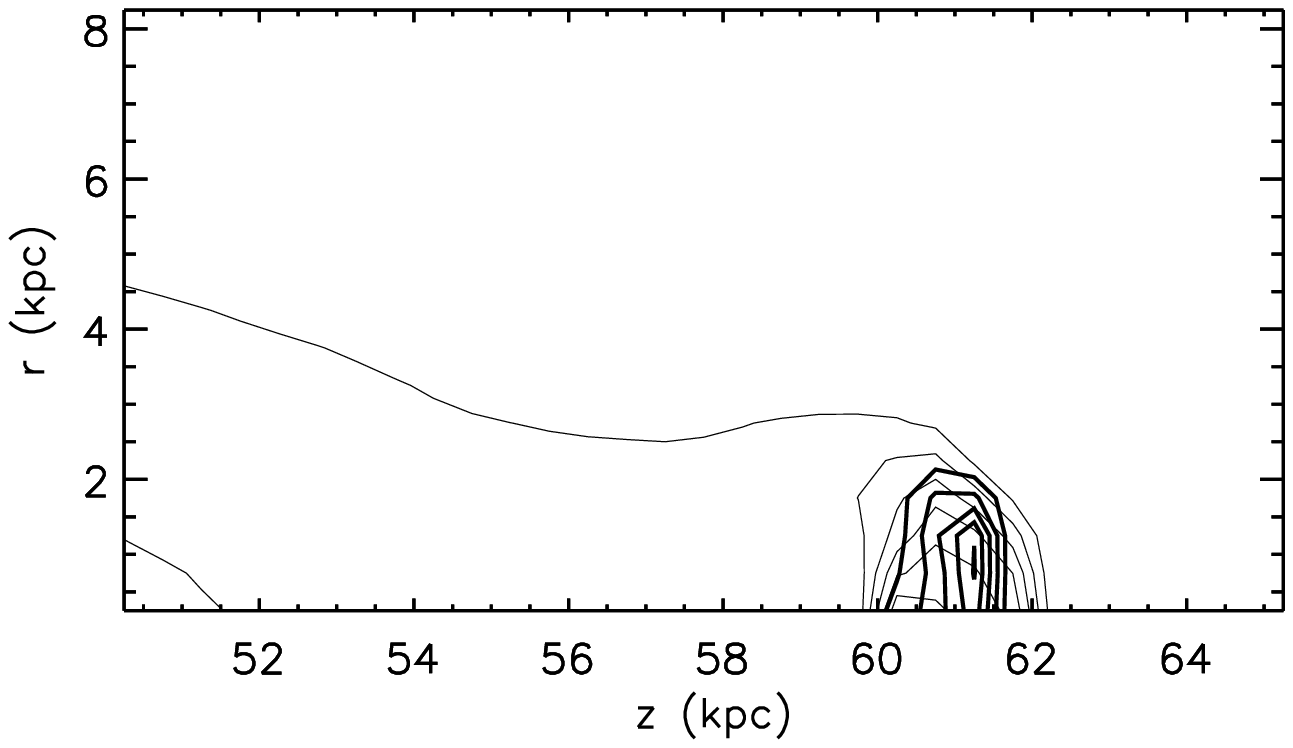}
\vskip.05in
\includegraphics[width=4.0in,scale=0.45,angle=0]
{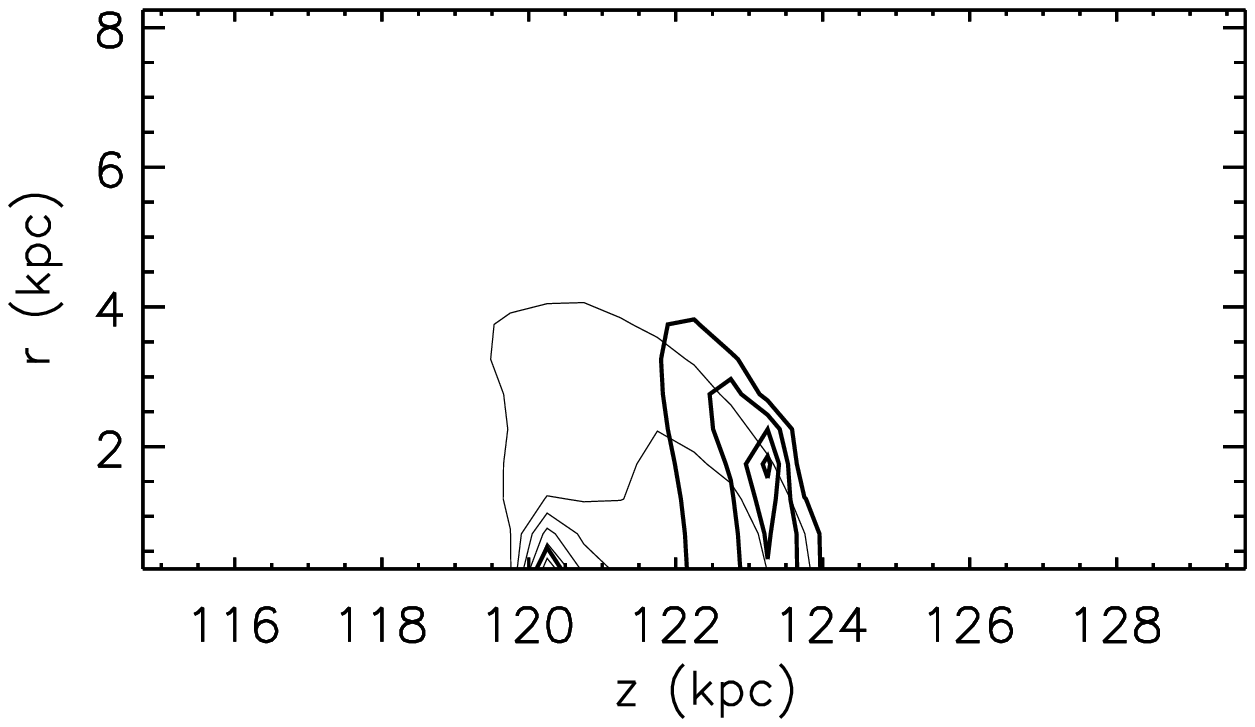}
\caption{
Radio synchrotron flux at 1.345 GHz ({\it heavy contours})  
and approximate surface brightness in X-ray IC-CMB radiation
({\it light contours}) for 
Cygnus A at its current age, 10 Myrs ({\it upper panel}) 
and at an age of 20 Myrs ({\it lower panel}).
}
\end{figure}


\clearpage

\begin{figure}
\centering
\includegraphics[width=4.0in,scale=0.45,angle=0]
{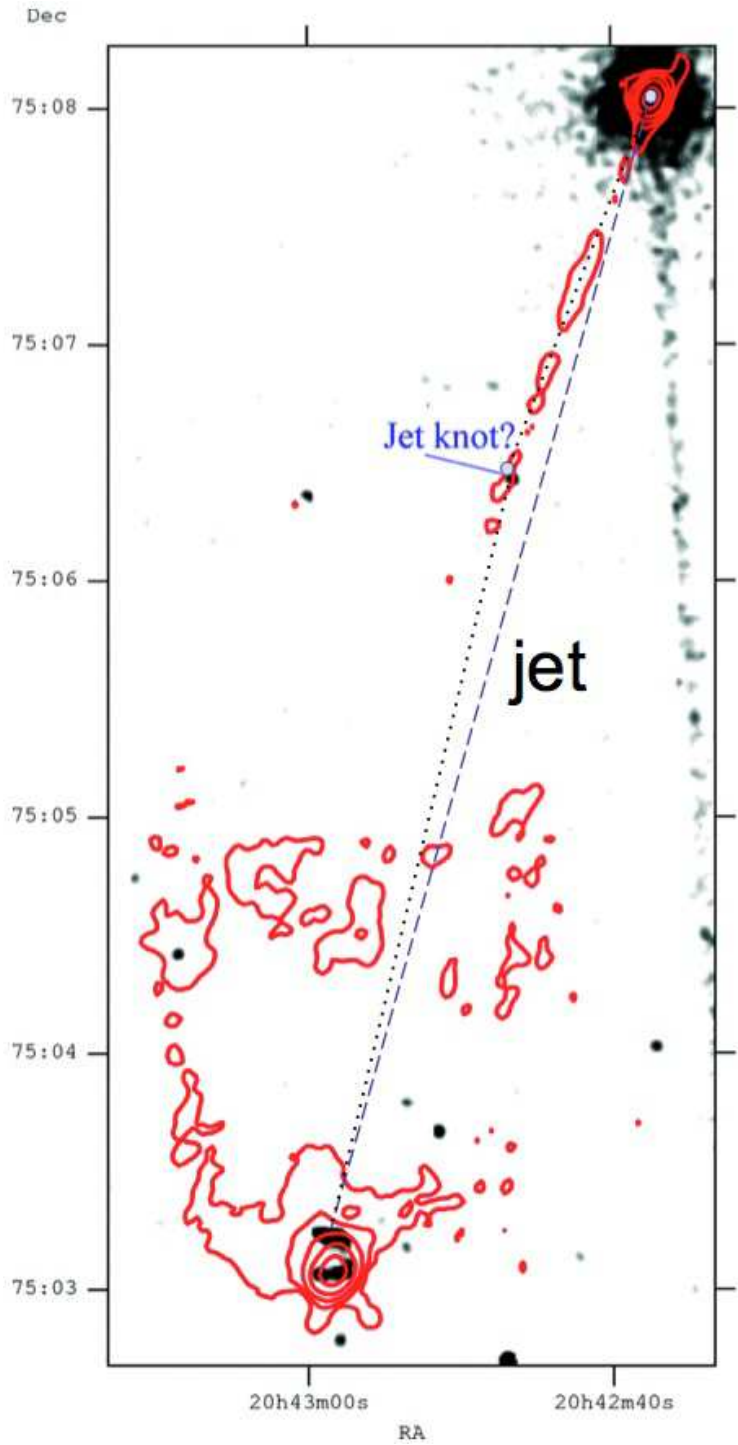}
\vskip.05in
\caption{
Radio emission at 1.4 GHz ({\it contours}) 
and X-ray emission ({\it gray scale}) from the FRII 
source 4C47.26 (Erlund et al. 2010). 
The shock spot and bright spot are both visible in X-rays.
The dashed and dotted lines are two possible jet trajectories  
that excited the currently observed shock spot.
}
\end{figure}

\end{document}